\documentclass[twocolumn, aps, showpacs,amsmath,superscriptaddress,notitlepage]{revtex4-2}
\usepackage{amssymb}
\usepackage{enumitem}

\usepackage[T1]{fontenc}
\usepackage{physics}
\usepackage{amsmath,amsfonts,bbm, graphicx, color,natbib}
\usepackage{amssymb}
\usepackage{appendix}
\usepackage{relsize}
\usepackage{xcolor}
\usepackage{scalerel}
\usepackage{tikz}

\usetikzlibrary{svg.path}
\usepackage[normalem]{ulem}
\usepackage{verbatim}
\usepackage{graphicx} 
\usepackage{subfigure}
\usepackage{bm}

\usepackage{alphalph}
\usepackage{etoolbox}
\usepackage{indentfirst} 
\usepackage{float}
\usepackage{xcolor}
\usepackage{booktabs}
\usepackage[normalem]{ulem} 
\usepackage{diagbox}
\usepackage{multirow}
\usepackage{makecell}
\usepackage{longtable} 
\usepackage{array}
\usepackage{tabularx}
\definecolor{orcidlogocol}{HTML}{A6CE39}
\def\blue{\textcolor{blue}}

\def\blue{\textcolor{blue}}

\def\blue{\textcolor{blue}}

\newcommand\orcid[1]{\href{https://orcid.org/#1}{\mbox{\scalerel*{
				\begin{tikzpicture}[yscale=-1,transform shape]
					\pic{orcidlogo};
				\end{tikzpicture}
			}{|}}}}
\usepackage[colorlinks]{hyperref}

\hypersetup{
	bookmarksnumbered,
	pdfstartview={FitH},
	citecolor={darkgreen},
	linkcolor={darkblue},
	urlcolor={darkblue},
	pdfpagemode={UseOutlines}}
\definecolor{darkgreen}{RGB}{20,100,20}
\definecolor{darkblue}{RGB}{0,0,130}
\definecolor{darkred}{rgb}{.8,0,0}

\usepackage{soul}

\makeatletter
\AtBeginDocument{\let\LS@rot\@undefined}
\makeatother

\begin{document}
	
	\def\qv{\vec{q}}
	\def\blue{\textcolor{blue}}
	\def\magenta{\textcolor{magenta}}
	\def\apricot{\textcolor{Apricot}}
	
	\def\GJ{\textcolor{black}}
	\def\TT{\textcolor{ForestGreen}}
	\definecolor{ora}{rgb}{1,0.45,0.2}
	\def\LH{\textcolor{black}}

\title{Non-Hermitian entanglement dip from scaling-induced exceptional criticality}
	\affiliation{School of Physics, Dalian University of Technology, Dalian 116024, China}
	\author{Sirui Liu}
	\affiliation{Department of Physics, National University of Singapore, Singapore 117551, Singapore}
	\affiliation{Department of Chemistry, National University of Singapore, Singapore 117543, Singapore}
	\affiliation{Joint School of National University of Singapore and Tianjin University,
			International Campus of Tianjin University, Binhai New City, Fuzhou 350207, China}
	\author{Hui Jiang}\email{phyjianghui@dlut.edu.cn}
		\affiliation{School of Physics, Dalian University of Technology, Dalian 116024, China}
	\affiliation{Department of Physics, National University of Singapore, Singapore 117551, Singapore}
	\author{Wen-Tan Xue}
	\affiliation{Department of Physics, National University of Singapore, Singapore 117551, Singapore}
	\author{Qingya Li}
	\affiliation{Department of Physics, National University of Singapore, Singapore 117551, Singapore}
	
		\author{Jiangbin Gong}
	\affiliation{Department of Physics, National University of Singapore, Singapore 117551, Singapore}
	\affiliation{Joint School of National University of Singapore and Tianjin University,
			International Campus of Tianjin University, Binhai New City, Fuzhou 350207, China}
			\affiliation{Centre for Quantum Technologies, National University of Singapore, 117543, Singapore}
	\author{Xiaogang Liu}
	\affiliation{Department of Chemistry, National University of Singapore, Singapore 117543, Singapore}
	\affiliation{Joint School of National University of Singapore and Tianjin University,
			International Campus of Tianjin University, Binhai New City, Fuzhou 350207, China}
		\author{Ching Hua Lee}\email{phylch@nus.edu.sg}

	\affiliation{Department of Physics, National University of Singapore, Singapore 117551, Singapore}
	\affiliation{Joint School of National University of Singapore and Tianjin University,
			International Campus of Tianjin University, Binhai New City, Fuzhou 350207, China}
		
	\date{\today}

\begin{abstract}
It is well established that the entanglement entropy of a critical system generally scales logarithmically with system size.
Yet, in this work, we report a new class of non-Hermitian critical transitions that exhibit dramatic divergent dips in their entanglement entropy scaling, strongly violating conventional logarithmic behavior. 
Dubbed scaling-induced exceptional criticality (SIEC), it transcends existing non-Hermitian mechanisms such as exceptional bound states and non-Hermitian skin effect (NHSE)-induced gap closures, which are nevertheless still governed by logarithmic entanglement scaling.
Key to SIEC is its strongly scale-dependent spectrum, where eigenbands exhibit an exceptional crossing only at a particular system size. 
As such, the critical behavior is dominated by how the generalized Brillouin zone (GBZ) sweeps through the exceptional crossing with increasing system size, and not just by the gap closure per se. 
We provide a general approach for constructing SIEC systems based on the non-local competition between heterogeneous NHSE pumping directions, and show how a scale-dependent GBZ can be analytically derived to excellent accuracy. 
Beyond 1D free fermions, SIEC is expected to occur more prevalently in higher-dimensional or even interacting systems, where antagonistic NHSE channels generically proliferate. 
SIEC-induced entanglement dips generalize straightforwardly to kinks in other entanglement measures such as Renyi entropy, and serve as spectacular demonstrations of how algebraic and geometric singularities in complex band structures manifest in quantum information.
\end{abstract}

	\maketitle

\noindent\textit{Introduction.--} 
Entanglement scaling behavior provides an established diagnostic for critical phase transitions. Deeply rooted in boundary conformal field theory~\cite{calabrese2004entanglement}, 
it universally associates logarithmic entanglement entropy (EE) scaling $S\sim \log L$ with 1D criticality~\cite{kitaev2006topological,kawabata2023entanglement}. This seminal scaling property has also seen generalizations to higher-dimensional and topological systems, where midgap modes and nodal structures contribute distinctive signatures to the entanglement scaling, some with measurable prospects.

Of late, established results in critical entanglement scaling have been challenged in non-Hermitian contexts. 
In so-called exceptional points (EPs)~\cite{PhysRevLett.123.066405}, geometric defectiveness blurs the distinction between occupied and unoccupied bands and probability non-conservation leads to fermionic occupancies effectively greater than one, giving rise to enigmatically negative free-fermion EE through the 2-point function~\cite{chang2020entanglement,lee2022exceptional}. This is further exacerbated by the non-Hermitian skin effect (NHSE), which produces macroscopically many highly defective eigenstates indexed by complex momenta. Yet, despite these complications, all known non-Hermitian free-fermion critical systems still exhibit the celebrated $S\sim \log L$ scaling.  

In this work, we unveil a peculiar new class of non-Hermitian critical transitions which departs dramatically from logarithmic entanglement scaling. Dubbed scaling-induced exceptional criticality (SIEC), it involves exceptional critical crossings that appear only at particular system sizes $L_c$, around which the entanglement entropy experiences characteristically divergent dips. 
As we shall elaborate through a rigorous generalized Brillouin zone (GBZ) construction, such dips can be traced to the breaking of scaling invariance which paradoxically lead to the emergence of supposedly scale-free gapless points.  \\


\begin{figure*}
	\centering
	\includegraphics[width=0.8\linewidth]{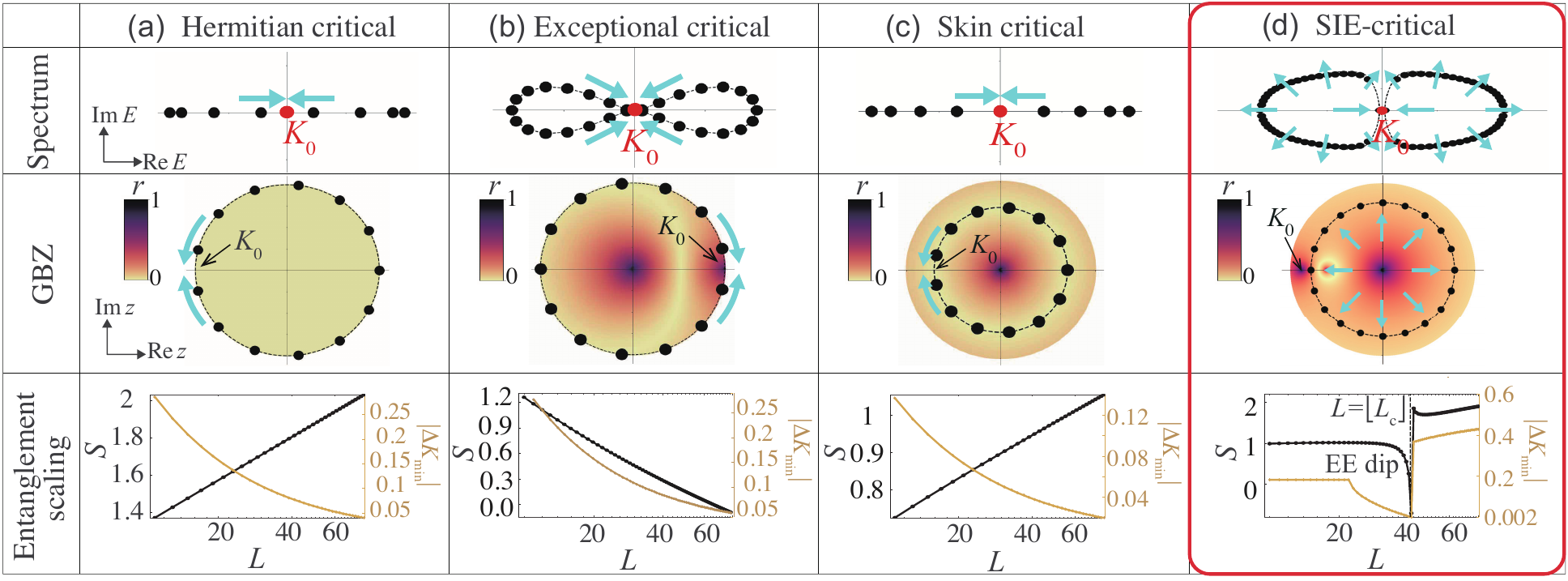}
\caption{\label{fig:fig1} Uniqueness of scaling-induced exceptional critical (SIEC) transitions in violating logarithmic entanglement scaling. (a) For a Hermitian critical point $K_0$, eigenmomentum states (black dots) simply become denser and approach $K_0$ (red) as the system size $L$ increases (cyan arrows), such that the closest approach $\Delta K_\text{min}\sim \pi/L$ (brown curve), giving rise to $S\sim \log L$ (black curve). 
(b) Conventional logarithmic EE scaling from usual $\Delta K_\text{min}\sim \pi/L$ eigenmomentum convergence still governs an exceptional non-Hermitian critical point $K_0$, albeit with a negative coefficient ($S\sim -\log L$) due to divergent 2-point functions at exceptional criticality where $r\rightarrow 0$ (dark purple).  
(c) Logarithmic EE scaling also persists under the NHSE, since we still have $\Delta K_\text{min} \sim \pi/L$  in the generalized Brillouin zone (GBZ)  $z=\text{e}^{iK (k)}$. 
(d) By contrast, in our SIE-critical (SIEC) systems, the spectrum and GBZ themselves change with $L$ (expanding cyan arrows), 
exhibiting critical touching $K_0$ (dark purple) only at a particular $L=\lfloor L_c\rfloor$. This leads to peculiar $\Delta K_\text{min}$ scaling and a characteristic entanglement dip in $S$.
}
\end{figure*}

\noindent\textit{Log-entanglement scaling from conventional non-Hermitian criticality.--} To appreciate the unconventional ingredients for our SIEC entanglement dips, we first review known types of critical scenarios. 
We write an arbitrary non-interacting 1D lattice Hamiltonian as 
\begin{equation}
\mathcal{H}=\sum_{k,\mu} E_\mu(k)|\psi_{k,\mu}^R\rangle\langle \psi^L_{k,\mu}|=\sum_{k,\mu} E_\mu(k)\psi_{k,\mu}^R[\psi^L_{k,\mu}]^*|k\rangle\langle k|,
\end{equation}
where $|k\rangle$ is the quasi-momentum basis state and $|\psi_{k,\mu}^{R}\rangle,\langle \psi^L_{k,\mu}|$ are respectively the right and left eigenbands~\cite{Brody_2014} of $\mathcal{H}$ with band index $\mu$. We stipulate that fermions occupy an arbitrary set of bands $\mu\in occ.$, such that state occupancy is described by the occupied band projector 
$P=\sum_{k,\mu\in occ.} |\psi_{k,\mu}^R\rangle\langle \psi^L_{k,\mu}|=\sum_kP_k|k\rangle\langle k|$, where $P_k=\sum_{\mu\in occ.}\psi_{k,\mu}^R[\psi^L_{k,\mu}]^*$. 
For an entanglement cut where the region $[x_L,x_R]$ is to be truncated, the biorthogonal free-fermion EE is given by~\cite{kawabata2023entanglement}
\begin{equation}
S= -\text{Tr}[\bar P\log\bar P +(\mathbb{I}-\bar P)\log(\mathbb{I}-\bar P)],
\label{S}
\end{equation}
where $\bar P = \bar{\mathcal{R}}P\bar{\mathcal{R}}$ is the occupied band projector that is restricted to $x\notin [x_L,x_R]$ by the real-space projector $\bar{\mathcal{R}}=\sum_{x\notin [x_L,x_R]}|x\rangle\langle x|$. In real-space, $\bar P$ is a block Toeplitz matrix with nonvanishing blocks $\bar P_{xx'}=\langle x | \bar P|x'\rangle$ (for $x,x'\notin [x_L,x_R]$) given by the Fourier transform 

\begin{eqnarray}
\bar P_{xx'}
&=& \sum_{k} \text{e}^{ik(x'-x)}\sum_{\mu\in occ.}\psi_{k,\mu}^R[\psi^L_{k,\mu}]^*=\sum_{k} \text{e}^{ik(x'-x)}P_k.\notag\\
\label{barP}
\end{eqnarray}


\noindent Based on  Eq.~\eqref{barP}, we first review why the EE scales like $\log L$ in Hermitian critical points
, as illustrated in Fig.~\ref{fig:fig1}a. Consider a  critical point $K_0$ (red) where the gap closes but the eigenspace remains full-rank i.e. non-defective. Due to the possible mixing of the occupied and unoccupied eigenstates at gap closure, 
the occupied band projector $P_k$ generically become discontinuous i.e. singular at the critical point. However, due to the discrete momentum spacing (black dots) which scales like $2\pi/L$ (cyan arrows) in a finite lattice, the momentum sum in  Eq.~\eqref{barP} only samples $P_k$ down to a distance $\Delta k =k-K_0$ bounded below by $\Delta K_\text{min}\sim\pi/L$ (brown curve) from the critical point i.e.  
$\bar P_{xx'}\sim\int_{\pi/L} \text{e}^{i\Delta k(x'-x)}P_{\Delta k} \,d(\Delta k)$ 
with a UV cutoff $\pi/L$. 
Since analytic singularities lead to power-law decay in their Fourier coefficients, we expect $P_{xx'}$, which are also the 2-point functions, to become long-ranged and cause substantial entanglement across the cut. Mathematically, this entanglement can be shown to diverge logarithmically with the inverse of the UV cutoff, thereby yielding $S\sim \log L$ entanglement scaling. 

Such $\log L$ entanglement scaling persists even if the critical point is geometrically defective (with completely non-orthogonal eigenstates), as is possible in non-Hermitian systems, as illustrated in Fig.~\ref{fig:fig1}b. For a particular band, the overlap between its left and right eigenstates $\langle\psi^L_k|$, $|\psi^R_k\rangle$ can be quantified through the phase rigidity~\cite{Ding2022}
$r(k)= \frac{\langle\psi^L_k|\psi^R_k\rangle}{\sqrt{\langle\psi^L_k|\psi^L_k\rangle\langle\psi^R_k|\psi^R_k\rangle}}, $
which is always unity in Hermitian systems, but ranging between $0\leq r(k)\leq 1$ in non-Hermitian systems. In particular, at an exceptional critical point (EP) $K_0$ where $r(K_0)=0$~\cite{PhysRevResearch.5.033042} (dark purple), the occupied and unoccupied eigenstates coalesce and the occupied band projector $P_k$ is not just discontinuous, but in fact divergent. Specifically, one or more matrix elements of $P_k$ can be shown to diverge with $\Delta k=k-K_0$ as $(\Delta k)^{-B}$, which scales with $\Delta K_\text{min}\sim L^{B}$, the model-dependent exponent $B$ related to the order of the EP~\cite{lee2022exceptional,ZOU2024}. 
Such divergences in $P_k$ have also been recently shown to lead to $S\sim \log L$ entanglement scaling, albeit with a negative coefficient determined by $B$ ~\cite{lee2022exceptional,ZOU2024}, as presented in the Supplementary material Section II.


Another key non-Hermitian phenomenon is the non-Hermitian skin effect (NHSE), but below we explain why the critical $\log L$ EE scaling still generically holds, as shown for the prototypical gapless non-Hermitian SSH [Fig.~\ref{fig:fig1}c] and other longer-ranged asymmetric hopping models. In general, NHSE "skin" states can be modeled with exponentially decaying state profiles $\text{e}^{ikx}\text{e}^{-\kappa(k)x}$ for each wavenumber $k$, the inverse decay length $\kappa(k)$ 
determined through detailed boundary condition analysis~\cite{yao2018edge, PhysRevLett.123.066404}. 
As such, the periodic and open boundary condition (PBC and OBC) eigenstates are approximately related through a complex quasi-momentum deformation $k\rightarrow K(k)=k+i\kappa(k)$, such that $\psi^R_{k,\mu}\rightarrow \psi^R_{K,\mu}$. 
This 
path $K(k)$, $k\in [0,2\pi)$ is known as the generalized Brillouin zone (GBZ), and can be visualized as a closed loop $z=\text{e}^{iK(k)}$ that deviates from the unit circle $\text{e}^{ik}$ due to the NHSE, as illustrated in Fig.~\ref{fig:fig1}c.

Importantly, the modified entanglement entropy due to the NHSE can be directly computed through this GBZ deformation, since the eigen-equations for the left and right eigenstates depend on $K(k)$ according to
\begin{align}
\mathcal{H}(K)|\psi^R_{K,\mu}\rangle&=E_\mu(K)|\psi^R_{K,\mu}\rangle\notag\\
\mathcal{H^\dagger}(K^*)|\psi^L_{K^*,\mu}\rangle&=E^*_\mu(K^*)|\psi^L_{ K^*,\mu}\rangle,
\label{biortho}
\end{align}
such that in the presence of skin state accumulation, $P_k\rightarrow \sum\limits_{\mu\in occ.}|\psi^R_{K,\mu}\rangle[|\psi^L_{ K^*,\mu}\rangle]^\dagger= P_K$. 
As such, replacing $P_k$ in  Eq.~\eqref{barP} by $P_K$, we obtain the 2-point functions under the NHSE as
\begin{eqnarray}
\bar P_{xx'} \xrightarrow[\,\text{NHSE}\,] \quad\sum_{k} \text{e}^{ik(x'-x)}P_{K(k)=k+i\kappa(k)}.
\label{barP2}
\end{eqnarray}
Note that the Fourier sum for $\bar P_{xx'}$ is still with respect to $k$ (and not $K$) because the NHSE does not alter the definition of the basis transform between real- and momentum-space.
If $P_{K}$ encounters a singularity $K_0$ along the GBZ path $K(k)$, then, like the above cases, the system would be critical and the EE would scale like 
\begin{equation}
|S|\sim |\log\Delta K_\text{min}|,
\label{SKmin}
\end{equation}
where $\Delta K_\text{min}$ is determined by the closest approach of $K(2\pi n/(L+1))$, $n=1,...,L$ to the singularity. Since $\Delta K_\text{min}$ would still scale like $L^{-1}$ (brown curve) except in rare cases with sufficiently badly behaved $K(k)$, the key takeaway is that the NHSE itself cannot give rise to critical entanglement scaling other than $S\sim \log L$. 

\noindent\textit{Unconventional criticality from scale-dependent GBZ.--} We have seen that two prominent non-Hermitian phenomena -- exceptional point (EP) defectiveness and the NHSE -- cannot by themselves lead to violations of $\log L$ entanglement scaling. This is because the closest approach to the critical point $K_0$ obeys $\Delta K_\text{min}\sim L^{-1}$ for all but the most pathological GBZ paths $K(k)$.  

In this work, a key insight is that qualitatively new entanglement scaling \emph{can} however occur if the GBZ trajectory $K(k)$ also depends on $L$ i.e. is scale-dependent. As $L$ increases, $\Delta K_\text{min}$ may vary non-monotonically with $L$, reaching a minimum when the $K(k)$ cuts across a singularity $K_0$, as highlighted in Fig.~\ref{fig:fig1}d in red. The GBZ loop $\text{e}^{iK(2\pi n/(L+1))}$, $n=1,...,L$ expands with increasing $L$ (outward cyan arrows on loop of black dots), such that it intersects $K_0$ (dark purple) only at a special value of $L=L_c$. As such, we have rapid unconventional $\Delta K_\text{min}\propto |L-L_c|$ scaling around $L\approx L_c$ (kink on brown curve). Since $S\sim -\log\Delta K_\text{min}$, we thus also expect $\log L$-violating EE scaling.

For such unconventional entanglement scaling, the system must meet the following conditions: (i) it must support qualitatively distinct dynamics at small $L$ and large $L$, such as to have very different spectra and hence GBZs in these limits; (ii) it must possess at least two bands (components) to exhibit an exceptional critical transition between these limits. 

\begin{figure}
	\centering
	\includegraphics[width=\linewidth]{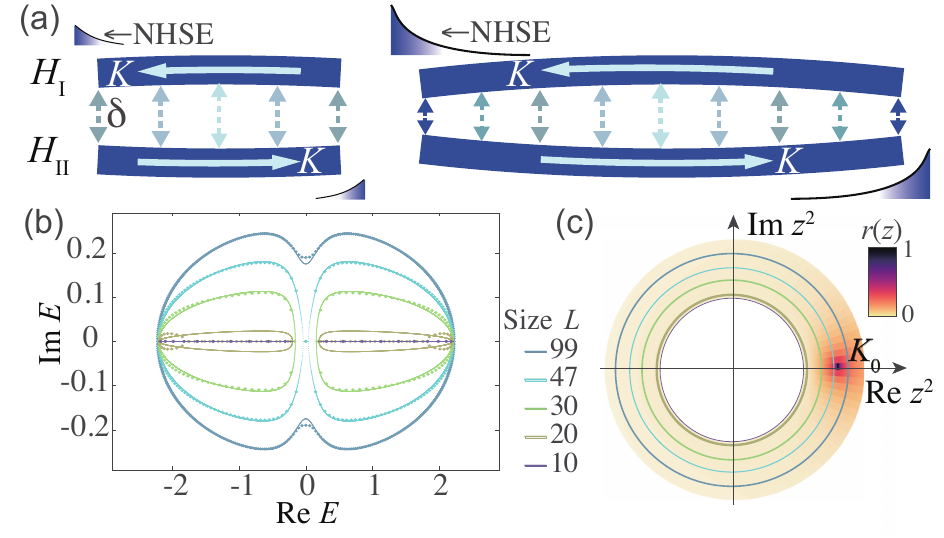}
	\caption{
Scaling-dependent critical spectrum and GBZ from antagonistic NHSE. 
(a) Our construction [ Eq.~\eqref{Hcoupled}] involves weakly coupling two chains $H$ with opposite NHSE and momentum label $k$. Only for sufficiently long chains (Right) is the NHSE strong enough to effectively close the two chains into a loop with PBC-like dynamics. 
(b)	Strongly $L$-dependent spectrum which undergoes a gapped to gapless exceptional critical transition at $L=47\approx L_c$ (cyan). (c) The corresponding $L$-dependent GBZ $z=\text{e}^{iK(k)}$ (colored loops), which pass through the critical point $K_0$ (where phase rigidity $r(z)=0$) when $L$ sweeps pass $L_c$. Parameters are $\delta = 8.4 \times 10^{-4}$ and $t_L=1.62$, $t_R=0.89$ in  Eq.~\eqref{SSH}.
	}
	\label{fig:fig2}
\end{figure}

Condition (i) can be generically met by weakly coupling two finite 1D NHSE chains $\nu=\text{I}$, $\text{II}$ with oppositely directed NHSE pumping to induce inter-chain tunneling [Fig.~\ref{fig:fig2}a]. As an important departure from usual literature, we label their basis $|x\rangle_\nu,|k\rangle_\nu$ in opposite directions [Fig.~\ref{fig:fig2}a], such that both chains are described by the same Hamiltonian $H$. This labeling, which assigns the momentum $k$ based on the NHSE pumping, would turn out instrumental in our GBZ construction. Our system, with weak coupling $\delta$, takes the form
\begin{eqnarray}
\label{Hcoupled}
\mathcal{H}
&=& \sum_{x,x'=1}^LH_{xx'}|x\rangle_\text{I}\langle x'|_\text{I}+\sum_{x,x'=1}^LH_{xx'}|x\rangle_{\text{II}}\langle x'|_{\text{II}}\notag\\
&& + \delta \sum_{x=1}^L \Big[|x\rangle_{\text{I}}\langle L+1-x|_{\text{II}}+|L+1-x\rangle_{\text{II}}\langle x|_{\text{I}}\Big]\notag\\
&=& \sum\limits_{k;\nu}H_{k}|k\rangle_\nu\langle k|_\nu+ \delta \sum_k \Big[|k\rangle_{\text{I}}\langle -k|_{\text{II}}+|-k\rangle_{\text{II}}\langle k|_{\text{I}}\Big],\qquad
\end{eqnarray}
where the last line describes only the bulk structure. Crucially, skin states grow exponentially with distance $L$ in each chain, concomitantly enhancing the effective inter-chain tunneling probability. At small $L$, we have two effectively uncoupled OBC chains (faint double-arrows in Fig.~\ref{fig:fig2}a). But at large $L$, the exponentially larger tunneling probabilities (dark blue double-arrows) effectively close the two antagonistic chains into a loop, forming a PBC-like configuration. 
This construction indeed produces a distinctively scale-dependent spectrum, as shown in Fig.~\ref{fig:fig2}b for a minimal illustrative model with coupled
\begin{equation}
H_\text{SSH}(k)=(t_L+\text{e}^{-ik})\sigma_++(t_R+\text{e}^{ik})\sigma_-
\label{SSH}
\end{equation}
chains, where $\sigma_\pm = (\sigma_x\pm i \sigma_y)/2$ are the Pauli matrices (see Fig.~\ref{fig:fig4} and the Supplementary material for other examples). 

While no exact analytic solution exists for the GBZ of generic $\mathcal{H}$ [ Eq.~\eqref{Hcoupled}], below we outline a general procedure for deriving an approximate GBZ that nevertheless captures its scaling properties accurately:
\begin{enumerate} [leftmargin =.5cm]
\item First, for an isolated chain $H$, solve $\text{Det}(H_k-E\,\mathbb{I})=0$ to obtain the bulk dispersion of $H$ i.e. relation between $z=\text{e}^{ik}$ and energy $E$. 

\item Solve the above dispersion polynomial and obtain all possible solutions $z_i$ for any given $E$. For each $z_i$, obtain the corresponding eigenstates $\ket{\zeta(z_i)}_\text{I}$ and $\ket{\zeta(z_i)}_\text{II}$ based on $H$ and $H^T$. While values of $E$ that satisfy the conventional GBZ condition $|z_i|=|z_j|$, $i\neq j$ correspond to the OBC spectrum of an \emph{isolated} chain, this condition will be severely violated for the coupled system $\mathcal{H}$, even with tiny $\delta \neq 0$.

\item Introduce the inter-chain coupling $\delta$ and solve the OBC eigenequation $\mathcal{H}\ket{\Psi_\text{GBZ}}=E\ket{\Psi_\text{GBZ}}$ for both coupled chains using the ansatz $\ket{\Psi_\text{GBZ}}=\sum_{i}c_i\sum\limits_{x=1}^L z_i^x\ket{\zeta(z_i)}_\text{I}\oplus z_i^{L+1-x}\ket{\zeta(z_i)}_\text{II}$, where $c_i$ are linear combination coefficients to be eliminated. The idea is to express the $\delta$ coupling terms entirely in terms of the OBC boundary contributions, both of which are what did not exist in the single-chain bulk solution in step 2 above. Although $\delta$ is very small, the coupling effect is non-perturbative due to the strong boundary sensitivity of non-Hermitian systems with the skin effect.

\item From the above, solve for the full GBZ solutions $z_i$ in terms of $L$, $\delta$ and $H$ model parameters, eliminating $E$ using the bulk dispersion from (1) if necessary. Due to the $z_i^{L+1-x}$ factor in our ansatz, which arises from the antagonistic momenta in  Eq.~\eqref{Hcoupled}, the GBZs $z_i$ typically satisfy a $L$-degree polynomial involving the model parameters, leading to generic $\sim 1/L$-th power scaling of $z_i$ with $\delta$. 

\end{enumerate}  

\noindent This GBZ effectively encapsulates the non-perturbative effects of the coupling $\delta$ through the scale-dependent complex deformation $K(k)$, henceforth allowing properties of the full system $\mathcal{H}$ to be computed just through a single chain $H$. As detailed in the Supplementary material and plotted in [Fig.~\ref{fig:fig2}c], the dominant scale-dependent GBZ for our minimal example $H_\text{SSH}$ [ Eq.~\eqref{SSH}] scales like
\begin{equation}
z_1=\text{e}^{iK(k)}\sim (\alpha_0\delta)^{1/(L+1)}\text{e}^{ik},
\label{z1}
\end{equation}
$\alpha_0$ a model-dependent constant.  Eq.~\eqref{z1} accurately predicts the spectrum of $\mathcal{H}(k)$  upon substituting $k\rightarrow -i\log z_1$ [Fig.~\ref{fig:fig2}b]. Correctly associating the antagonistic directions with opposite momenta [ Eq.~\eqref{Hcoupled}] was crucial in deriving this scaling-dependent GBZ and spectrum, which cannot be analytically derived as a function of $L$ using conventional approaches~\cite{li2020critical}.

\begin{figure}
	\centering
	\includegraphics[width=1\linewidth]{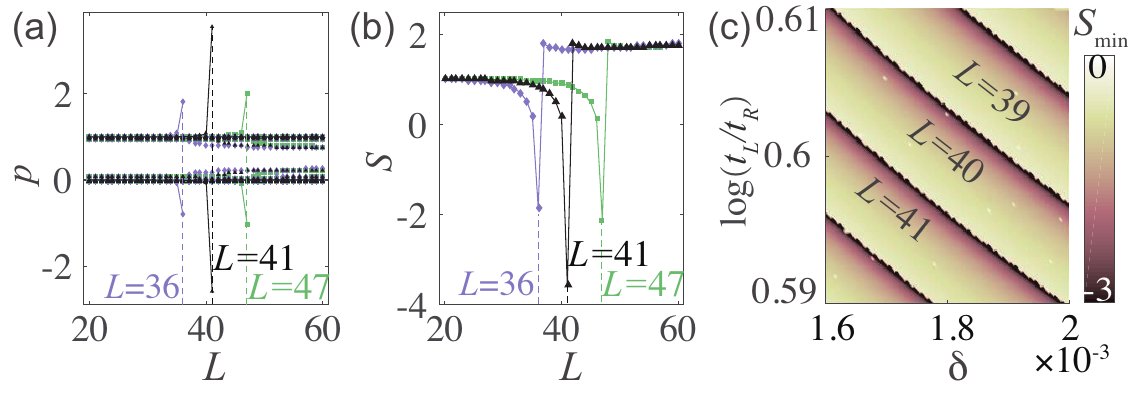}
	\caption{
	Entanglement dip from SIEC. (a) Spikes in the $\bar P$ eigenvalues $p$ occur near $L\approx L_c$ [ Eq.~\eqref{eqLc}] for different inter-chain couplings $\delta=0.83\times10^{-3}, 1.6843\times 10^{-3}$ and $3.03\times 10^{-3}$ (green, black and purple respectively) of  Eq.~\eqref{SSH}, with hopping asymmetries $t_L,t_R=1.2e^{\pm 0.3}$. 
	(b) Corresponding entanglement dips at these $L\approx L_c$, with characteristic $S\sim (L_c-L)^{-1/2}$ behavior for $L<L_c$.
	(c) Depth of entanglement dips in the parameter space of $t_L/t_R, \delta$ and the system sizes $L$ realizing them, with large dips of $S_\text{min}\leq -3$ along narrow bands separating regions of different $L$.}
	\label{fig:fig3}
\end{figure}

\noindent\textit{SIEC and entanglement dip.--} 
Our $L$-dependent GBZ can give rise to a new phenomenon dubbed ``scaling-induced exceptional criticality'' (SIEC) if the GBZ loop $\text{e}^{iK(k)}$ sweeps through an EP (where $r(K)=0$, dark purple in Fig.~\ref{fig:fig2}c) when $L$ is varied across a special value $L_c$. 
For our illustrative $H_\text{SSH}$ and related models, substituting  Eq.~\eqref{z1} into $H_\text{SSH}(K(k))$ and demanding that it reduces to the Jordan form $\sigma_-$ at the EP yields 
\begin{eqnarray}
\label{eqLc}
L_c&=&-\alpha_0\delta/\log t_R -1,\\ 
K_c&=&K|_{L=L_c} = \pi+\frac{i\alpha_0\delta}{L+1},
\label{eqKc}
\end{eqnarray}
which accurately predicts when the exceptional transition occurs, corroborated by numerical results in Fig.~\ref{fig:fig2}b. As $L_c$ is typically non-integer, actual lattice $\mathcal{H}$ models with integer $L$ experience dramatic SIEC divergences when $L\rightarrow L_c$, as controlled by
\begin{eqnarray}
\Delta_K=K(k)-K_c
&\approx & (k-\pi)+i\delta\frac{\alpha_0 (L_c-L)}{L_c^2},
\label{DeltaK}
\end{eqnarray}
such that  $H_\text{SSH}(K)\approx (t_L-t_R^{-1})\sigma_+-it_R\Delta_K\sigma_-$ for $L\approx L_c$, with the expected square root cusp in its eigenenergies $\pm\mathcal{E}_\text{SSH}(K)=\pm \sqrt{i(1-t_Lt_R)\Delta_K}$.
In particular, its occupied band projector, which assumes the form $P(K)=(\mathbb{I}-H(K)/\mathcal{E}(K))/2$ for 2-component models, has divergent off-diagonal term $-\sqrt{i(t_L-t_R^{-1})/4t_R\Delta_K}\,\sigma_+$. Hence, with $k=\pi$
, we obtain the unique scaling divergence
\begin{equation}
P\sim -\sqrt{\frac{t_L-1/t_R}{4t_R\alpha_0\delta }}\frac{L_c}{\sqrt{L_c-L}}\,\sigma_+ \ ,
\end{equation} 
which, unlike in conventional critical scenarios of Figs.~\ref{fig:fig1}b,c, diverges due to $L\rightarrow L_c$ rather than vanishing momentum spacings. This square-root divergence shows up in the $\bar P$ spectrum near $L\approx L_c$ [Fig.~\ref{fig:fig3}a, also see the Supplementary material] which in turn translates to a characteristic $S\sim 1/\sqrt{|L_c-L|}$ divergence in the EE which we call the \emph{entanglement dip} [Fig.~\ref{fig:fig3}b]. 

While such entanglement dips can in principle be infinitely deep, in practice its depth is limited by how closely $L_c$ approaches an integer. Shown in Fig.~\ref{fig:fig3}c is the minimal entanglement $S_\text{min}$ as the hopping asymmetry $t_L/t_R$ and inter-chain coupling $\delta$ are varied. For each lattice size $L$, the EE dip becomes particularly deep along curves in parameter space (dark pink), even reaching $S_\text{min}\approx -3$ in the region shown. \\




\begin{figure}
	\centering
	\includegraphics[width=1\linewidth]{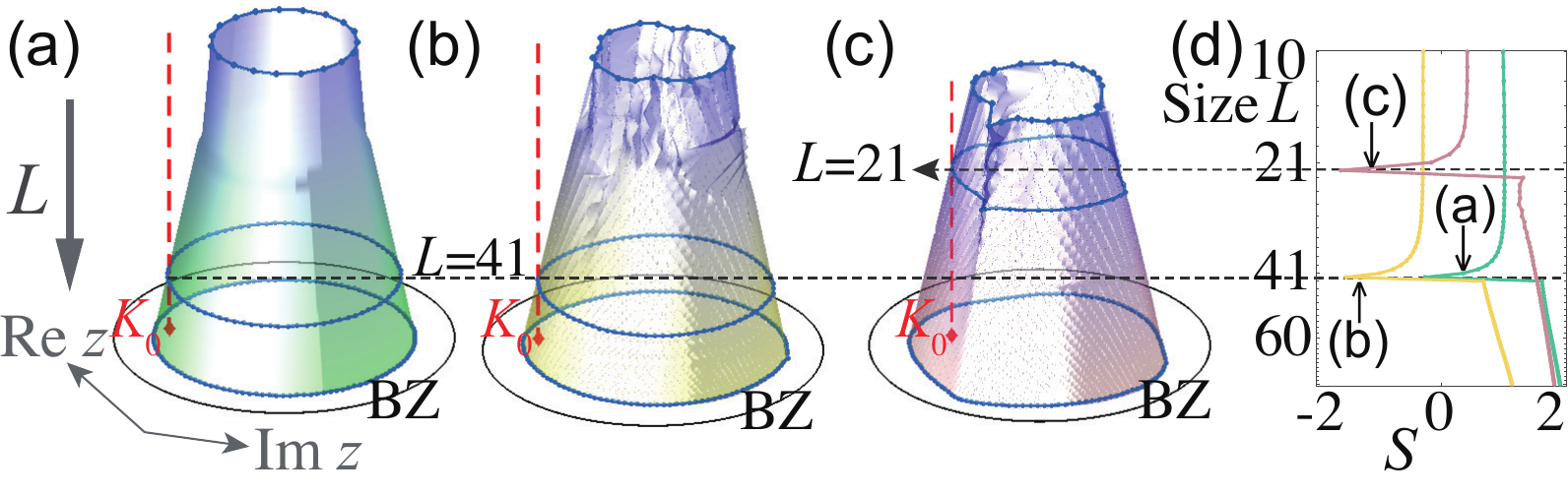}
	\caption{
Entanglement dips and scale-dependent GBZ in more general models satisfying  Eq.~\eqref{Hcoupled}. Shown are the $L$-dependent GBZs for illustrative (a) $H=(t_L+1/z)\sigma_+ + (t_R+z)\sigma_-$, (b)	$H=(t_L+1/3z^{-2})\sigma_+ + (t_R+z)\sigma_-$ (middle) and (c)	$H=t_L\sigma_+ + (0.6+0.12i+z+0.1z^{-2})\sigma_-$ (right), all with weak inter-chain coupling $\delta=1.6\times 10^{-3}$ and $t_L,t_R=1.2\text{e}^{\pm 0.3}$ (here, $L$ increases from top to bottom). 
(d) While their GBZs can be irregularly-shaped, entanglement dips in $S$ consistently emerge at values of $L$ where the GBZ loop intersects the critical point $K_0$ (red).
}
\label{fig:fig4}
\end{figure}

\noindent{\it Generality of entanglement dips.--} SIEC and entanglement dips are not limited to the coupled $H_\text{SSH}$ chains explicitly computed so far. They are expected to show up whenever the spectrum depends on $L$, and becomes gapless at an EP at a special $L\approx L_c$. This requires a scale-dependent GBZ that can be generically designed by coupling subsystems with competing NHSE pumping. 
Presented in Fig.~\ref{fig:fig4} are other models containing oppositely-directed NHSE channels, albeit with lower symmetry. Evidently, they all exhibit scale-dependent GBZs, and importantly exhibit entanglement dips at $L\approx L_c$ whenever the GBZ encounters a critical point $K_0$ (red). \\

\noindent{\it Discussion.--} 	
We have uncovered a new class of critical transitions marked by characteristic $S\sim (L_c-L)^{-1/2}$ EE dips, departing from the almost-universal $S\sim\log L$ critical scaling for free fermions, Hermitian or otherwise.  
Its peculiar scaling behavior originates from the scale-dependence of the GBZ itself, which fundamentally alters how a critical point can be approached. 

Physically, such dramatically suppressed EE represent the unique non-conservation of probability from antagonistic non-Hermitian pumping, which never occurs in ordinary NHSE processes where the gain/loss can be "gauged away" with a basis redefinition. Beyond the EE, entanglement dips also translate to kinks in the Renyi entropy, whose measurement prospects we discuss in Supplementary material Section V.\\



\noindent{\it Conflict of interest.} The authors declare that they have no conflict of interest.

\noindent{\it Acknowledgements.}  J.G. acknowledges support by the National Research Foundation, Singapore and A*STAR under its CQT Bridging Grant. C.H.L acknowledges support from the Ministry of Education, Singapore (MOE award number: MOE-T2EP50222-0003). H.J  is supported by “the Fundamental Research Funds for the Central Universities.


\noindent{\it Author contributions.} Ching Hua Lee proposed the initial idea and led the whole project. Hui Jiang co-led the project and designed key calculations. Sirui Liu and Hui Jiang executed the project and conducted key calculations. Wen-Tan Xue and Qingya Li performed
numerical simulations and data analysis. Jiangbin Gong and Xiaogang Liu participated in the discussions and supervised the research. All participated in the writing of the manuscript.


\bigskip
\clearpage

\onecolumngrid
\begin{center}
	\textbf{\large Supplementary Materials}\end{center}
\setcounter{equation}{0}
\setcounter{figure}{0}
\renewcommand{\theequation}{S\arabic{equation}}
\renewcommand{\thefigure}{S\arabic{figure}}

\section{How to obtain the system size-dependent effective GBZ and spectrum} \label{suppmat}
Recent work has revealed the crucial importance of the generalized Brillouin zones (GBZ) in describing non-Hermitian lattice Hamiltonians open boundary conditions (OBCs)~\cite{PhysRevLett.125.126402,Mong2011winding,slager2015impurity,PhysRevB.102.205118,zhang2021acoustic}. For a given Hamiltonian $H(z)$ where 
 $z=\exp(ik)$, the two roots $z$ of the characteristic (dispersion) equation $\text{det}(H(z)-E)=0$ with equal absolute values $|z|$ represents two ``standing wave'' solutions that can be superposed to satisfy OBCs at both ends. This is the typical way the GBZ is defined~\cite{yao2018edge,xiong2018does,PhysRevB.99.201103,song2019realspace,longhi2019probing,PhysRevLett.125.126402,PhysRevLett.123.066404,PhysRevLett.125.226402,lee2020ultrafast,longhi2020non,lee2020unraveling}.

However, there exist many systems where the usual GBZ approach gives a poor approximation to the actual numerical spectrum. Consider two weakly coupled chains exhibiting the non-Hermitian skin effect (NHSE) in opposite directions. In the weak coupling limit, we expect the presence of localized skin states at the ends of the individual chains. Due to the oppositely directed skin effect, these state accumulations must also be oppositely directed. Consequently, it is not tenable to exploit the same ``standing waves''  to satisfy the system's boundary conditions.

In other words, for such systems with competitive skin effect channels, it is not possible to accurately model the the net skin effect using the usual GBZ approach. Below, we shall present a new alternative approach using \emph{different} momentum wavevector directions in different chains, and show that how we can derive a \emph{size-dependent GBZ} which accurately considers the effect of the coupling different skin effect channels.



\subsection{ Coupled 1D chains with opposite NHSE}\label{suppmat1a}

We write down a model in which non-Hermitian 1D chains with antagonistic NHSE, denoted by $H$ and $H^{T}$, are weakly coupled together by a small coupling parameter $\delta$. In the simplest cases, the coupling acts only between corresponding pairs of sites in a ladder-like fashion, such that the coupled Hamiltonian takes the form
\begin{equation}\label{EqS1}
	\begin{split}
		&	\mathcal{H} =
		\begin{pmatrix} 
			H & \delta \mathbb{I}  \\
			\delta  \mathbb{I} & H^{T}
		\end{pmatrix}\ ,\\
	\end{split}
\end{equation}
where $\delta \mathbb{I}$ is the identity matrix acting in the basis of $H$ or $H^T$, as shown in Fig.~\ref{picS1}a.  We specialize $H$ to the well-known non-Hermitian SSH model~\cite{yao2018edge,yin2018geometrical}, which is simple enough for analytic solutions and yet topologically nontrivial:
\begin{equation}\label{EqS2}
	\begin{split}
		H =\sum_n \begin{pmatrix}
			0 & 1\\
			0& 0\\
		\end{pmatrix}|n\rangle\langle n-1|+\begin{pmatrix}
			0 & t_L\\
			t_R& 0\\
		\end{pmatrix}|n\rangle\langle n|+ \begin{pmatrix}
			0 & 0\\
			1& 0\\
		\end{pmatrix}|n\rangle\langle n+1|\ .
	\end{split}
\end{equation} 
By diagonalizing $H$ in the Fourier basis, we obtain eigenenergies
\begin{equation}
 {E}=\pm \sqrt{(t_L + 1/z )(t_R + z )}\ .
\label{Ez}
\end{equation}
Under periodic boundary conditions (PBCs), we simply have $z=\text{e}^{ik}$ where $k$ is the momentum.
In the absence of any NHSE, this should also approximately represent ``bulk'' spectrum under open boundary conditions (OBCs), since boundary effects should only affect a subdominant proportional of the eigenstates.

However, it is well-known that due to the NHSE, the boundaries would significantly affect the \emph{entire} spectrum, such that we must have~\cite{yao2018edge} $z = \text{e}^{ik - \kappa(k)}$ for $E(z)$ to correctly approximate the OBC spectrum. Here, $\kappa(k)$ represents the complex deformation of the momentum $k \rightarrow k + i\kappa(k,\kappa \in \mathbb{R})$, which is what defines the GBZ. 

Here, we first show how to compute the GBZ of fully \emph{uncoupled} $H$ chains. We note that to satisfy OBCs at both ends of a 1D chain, an eigenfunction$|\Psi\rangle$ must be a linear combination~\cite{yao2018edge}  of at least two different non-Bloch wave functions {$\sum_n z_1^n( |n,A\rangle+\phi_{z_1}|n,B\rangle)$ and $\sum_nz_2^n ( |n,A\rangle+\phi_{z_2}|n,B\rangle)$, where $A,B$ and $n$ label the sublattices and unit-cells, $z_1,z_2$ are the two solutions of Eq.~\eqref{Ez} which satisfy $z_1z_2=t_R/t_L$, and $\phi_{z_{1,2}}=(t_R+z_{1,2})/E$. Crucially, the OBCs $\langle 0,B|\Psi\rangle=\langle L+1,A|\Psi\rangle=0$ enforce the constraint $z_1^{L+1}\phi_{z_2}=z_2^{L+1}\phi_{z_1}$ such that $(\phi_{z_1}/\phi_{z_2})^{1/L}\approx1$.} Consequently, for our uncoupled chain $H$, the GBZ satisfies $|z_1|=|z_2|$ and is simply given by $\left\{z=\sqrt{{t_R}/{t_L}}\exp(ik)|\,k=\frac{{2}m\pi}{L+1}, m=1,2,\cdots, L\right\}$ i.e. with skin depth  $\kappa_{H}=\frac1{2}\log |z|=\frac{1}{2}\log(t_R/t_L)$. Its  {transpose} conjugate is an oppositely directed chain $H^{T}$ that possesses skin depth $\kappa_{H^{\dagger}}=-\kappa_{H}=-\frac{1}{2}\log(t_R/t_L)$.

Note that the condition $|z_1|=|z_2|$, which is commonly lauded as the GBZ constraint, will no longer be valid once we consider coupled chains with different NHSE directions.
\\

Before we discuss the coupled ($\delta\neq 0$) case, we note that our system possesses spatial inversion symmetry $\mathcal{I}$ represented by 
\begin{eqnarray}
	\mathcal{I} \mathcal{H}\mathcal{I}=\mathcal{H},\qquad \mathcal{I}=\sum_n(\tau_x\otimes \sigma_x)\ket{L+1-n}\bra{n},
\end{eqnarray}
such that the wave function accumulates similarly at both ends of the chain. This $\mathcal{I}$ symmetry is a consequence of the inversion symmetry between $H$ and $H^\dagger$, and is \emph{not} fundamental to the general construction of our size-dependent effective GBZ. However, it offers to simplify the system's boundary conditions by half, and is thus helpful to obtaining the analytic results that follow.

\subsection{  The new approach:  {Scale}-dependent energies and GBZ}\label{suppmat1b}
We next derive the size-dependent {effective} GBZ for two antagonistic NHSE chains coupled by a non-zero but small $\delta$ coupling. Since $\delta$ is small, the effect of the coupling term is insufficient to alter the orientations of the NHSE within each of the two chains, which are in opposite directions. As such, the usual GBZ approach, which assigns the \emph{same} decay parameter  {magnitude $|z|$} to all subsystems (i.e. both chains), is doomed to be inadequate. 

Instead, in our approach, we assign the decay parameter $z$ to one chain and $ {1/z}$ 
to the other chain.  {Ultimately, the OBC eigenstate would in general consist of a linear combination of two or more eigensolutions, which we will first solve for, separately. Writing each $z$ eigensolution as $ \ket{\varphi}=\sum_{n,\alpha}\varphi_{\alpha,n}\ket{n,\alpha}$} with sublattice $\alpha= A, B$(chain I), $A', B'$(chain II), 
   {spatial inversion symmetry $\mathcal{I}$ enforces}  {$\varphi_{A(B),n}=\varphi_{B'(A'), L+1-n}$. Concretely,
\begin{eqnarray}
&\varphi_{A,n}=z^n,&\ \varphi_{B,n}=\phi_z z^n\ ;\notag\\
&\varphi_{A',n}=\phi_z z^{L+1-n},&\  \varphi_{B',n}=z^{L+1-n}\ .
\label{psiAB}
\end{eqnarray}
Due to spatial inversion symmetry $ {\mathcal{I}}$, the decay in both chains are exactly equal but opposite -- in more generic cases where the 2 coupled chains are not related by spatial inversion symmetry $ {\mathcal{I}}$, they would need to be separately solved for. }

In the above, the assignment of $z$ and $1/z$ decay parameters to chains I and II only addresses their relative decay directions. To solve for what value/s the decay parameter $z$ should take, we need to incorporate the weak couplings and the open boundary conditions. In our approach, the idea is to (i) first solve for the bulk relation between $z$ and the corresponding eigenenergy $E$ in the uncoupled limit, and then (ii) invoke the weak couplings and boundary conditions to obtain a size-dependent effective GBZ consistent with that relation.

In our given Hamiltonian $\mathcal{H}$, the full bulk equations take the form
\begin{eqnarray}
	&&\left\{\begin{array}{ll}
		t_L {\varphi}_{B,n} +      {\varphi}_{B,n-1} + \delta   {\varphi}_{A',n} &= E  {\varphi}_{A,n}\\
		t_R {\varphi}_{A,n} +       {\varphi}_{A,n+1} + \delta   {\varphi}_{B',n} &= E  {\varphi}_{B,n}\\
		t_R {\varphi}_{B',n} +      {\varphi}_{B',n-1} + \delta   {\varphi}_{A,n}& = E  {\varphi}_{A',n}\\
		t_L {\varphi}_{A',n} +       {\varphi}_{A',n+1} + \delta   {\varphi}_{B,n} &= E  {\varphi}_{B',n}
	\end{array}\right. \quad \stackrel{ {\mathcal{I}}}{\longrightarrow} \left\{\begin{array}{ll}
		t_L {\varphi}_{B,n}+      {\varphi}_{B,n-1}  + \delta   {\varphi}_{B,L+1-n} &= E  {\varphi}_{A,n}\\
		t_R {\varphi}_{A,n} +       {\varphi}_{A,n+1} + \delta   {\varphi}_{A,L+1-n} &= E  {\varphi}_{B,n}\\
	\end{array}\right. ,
\end{eqnarray}
with site $n=2,3,\cdots,L-1$ and eigenvalue $E$. Due to $ {\mathcal{I}}$ symmetry, the four equations (from two chains with two sublattices each) are simplified to two equations. Substituting Eq.~\eqref{psiAB} into the above, 
\begin{eqnarray}
	\left\{\begin{array}{rl}
		(-E+(t_L+     1/z)\phi_z)z^n+ \phi_z z^{L+1-n}\delta&=0\\
		((t_R+     z) -E\phi_z)z^n+  z^{L+1-n}\delta &=0
	\end{array}\right.  \ .
\end{eqnarray}
For step (i), we work in the $\delta\ll 1$ limit and neglect the inter-chain couplings $\delta$.
 In practice, it is sufficient for $\delta$ to be at least one order of magnitude smaller than other system parameters. This yields the decoupled eigenequation to obtain 
\begin{eqnarray}
	\left\{\begin{array}{rl}
		(-E+(t_L+     1/z)\phi_z)z^n&\approx 0\\
		((t_R+     z) -E\phi_z)z^n&\approx 0
	\end{array}\right.  \Rightarrow\begin{pmatrix} 
		-E& t_L+    1 /z\\
		t_R+     z& -E
	\end{pmatrix}\begin{pmatrix} 
		1 \\
		\phi_z
	\end{pmatrix}=(H(z)-E)\begin{pmatrix} 
		1 \\
		\phi_z
	\end{pmatrix} =0 \  .
\end{eqnarray}
Via the same steps as in the previous subsection on a single chain, these bulk equations give (Eq.~\eqref{Ez})
\begin{eqnarray}
	E=\pm \sqrt{(t_L+     1/z)(t_R+     z)}\ ,\quad \phi_z=\frac{t_R+     z}{E} =\frac{E}{t_L+  1/ z}\ ,
	\label{Eqenergy}
\end{eqnarray}
which, for a particular eigenvenergy $E$, corresponds to two 2 solutions $z=z_1,z_2$ which satisfy 
\begin{align}
	&\qquad E^2 = (t_L+ 1/z)(t_R+ z)\  , \\
	&t_Lz^2 + (t_Rt_L + 1 - E^2)z + t_R = 0\ .
\end{align}
WLOG, we label them in the order $|z_2|\leq \sqrt{|t_R/t_L|}\leq |z_1|$ since, according to Vieta's formulas,
\begin{eqnarray}
	z_1z_2= t_R/t_L\ .
	\label{z1z2}
\end{eqnarray}
 {Had the 2-component subsystem $H(z)$ existed in isolation, the GBZ solution would just have been given by $|z_1|=|z_2|$, with the OBC skin spectrum given by the set of $E$ satisfying this equality constraint. However, in the presence of (even very weak) inter-chain couplings, the incorporation of the full boundary conditions will give rise to a very different constraint from $|z_1|=|z_2|$, as we will soon see.}



We next proceed to step (ii) where we 
consider the boundaries at unit-cells $1$ and $L$ without neglecting the weak couplings $\delta$.  {To satisfy the boundary conditions, we now have to consider an eigensolution that superposes all the (two) different single-chain GBZ solutions $z_1,z_2$ from step (i), we write the ansatz OBC eigenfunction as}
\begin{eqnarray}\label{Eqpsi}
	\ket{\Psi_{\text{GBZ}}}= {|\varphi_{z_1}\rangle-c|\varphi_{z_2}\rangle}=\sum^L_{n=1}\sum_{\alpha=A,B,A',B'}\psi_{n,\alpha} |n,\alpha\rangle=\sum_n \left(\begin{pmatrix} 
		z_1^n\\
		z_1^n\phi_{z_1}\\
		z_1^{L+1-n}\phi_{z_1}\\
		z_1^{L+1-n}
	\end{pmatrix}-c \begin{pmatrix} 
		z_2^n\\
		z_2^n\phi_{z_2}\\
		z_2^{L+1-n}\phi_{z_2}\\
		z_2^{L+1-n}
	\end{pmatrix}\right)\ket{n},
\end{eqnarray}
 {where the coefficient $c$ controls the relative amplitude of the $z_1,z_2$ non-Bloch contributions.} We next substitute this ansatz Eq.~\eqref{Eqpsi} into the open boundary conditions, where $\psi_{n,\alpha}$ disappear at $n=0$ and $L+1$ (and beyond):
\begin{eqnarray}
\label{eqS13}
	&&\left\{\begin{array}{ll}
		t_L\psi_{B,1} + \delta  \psi_{A',1}  &= E \psi_{A,1}\\
		t_R\psi_{A,L} + \delta  \psi_{B',L}  &= E \psi_{B,L}\\
		t_R\psi_{B',1}  + \delta  \psi_{A,1} & = E \psi_{A',1}\\
		t_L\psi_{A',L}  + \delta  \psi_{B,L} &= E \psi_{B',L}
	\end{array}\right. \stackrel{ {\mathcal{I}}}{\longrightarrow}\left\{\begin{array}{ll}
	 {\delta\psi_{B,L}= E\psi_{A,1}-t_L\psi_{B,1}} &\\
	 {\delta  \psi_{A,1}=E\psi_{B,L}-t_R\psi_{A,L}} &\\
			\end{array}\right..\end{eqnarray}
 {To proceed, the \emph{key} idea is that the extra terms involving the small inter-chain coupling $\delta$, which were not taken into account in deriving the bulk $z_1,z_2$ solutions, must exactly compensate the missing terms due to the open boundaries. By explicitly substituting $E$ in terms of $z_{1,2}$, $\phi_{z_{1,2}}$ and the hoppings using Eq.~\eqref{Eqenergy}, we obtain}
\begin{eqnarray}
	&&\left\{\begin{array}{rl}
		(  1 - \delta z_1^L)\phi_{z_1}&= c(    1 - \delta z_2^L)\phi_{z_2}\\
		(  z_1^{L}- \delta)z_1&= c( z_2^{L}- \delta)z_2
	\end{array}\right.\ , 
\end{eqnarray}
which yields
\begin{eqnarray}
	(      1- \delta z_1^L)(       z_2^{L}- \delta)z_2\phi_{z_1}=(  1   - \delta z_2^L) (       z_1^{L}- \delta)z_1 \phi_{z_2}\ .
\end{eqnarray}
 {We comment that $z_1,z_2$, which were previously derived under the uncoupled ($\delta=0$) approximation, has now become dependent on the coupling $\delta$ through $E$ (even though this dependence is hidden behind Eq.~\eqref{Eqenergy}, which was derived under the uncoupled approximation).
} 
Since the inter-chain couplings are weak  {and the spatial inversion symmetry}, we are safe to assume that they cannot fundamentally reverse the NHSE direction within each chain. That is, in the case of $|t_R/t_L|<1$, 
we have $ |z_2| \leq \sqrt{|t_R/t_L|} \leq |z_1|\leq 1$. 
Together with $\delta\ll 1$ and $|z_{j}|<1$($j=1,2$),    {note that for small systems $|z_j|$ is well below 1, whereas for large systems $|z_j|$  may approach 1 yet  $\lim_{L\gg1}|z_j|^L$ still decays rapidly to zero. Moreover, the imaginary part of  $z_j^L$ can be neglected since the effective model requires the momentum to satisfy periodic boundary conditions. Therefore, we can approximate} $1 - \delta z_{j}^L\approx 1$ ($j=1,2$), thereby simplifying the above to
\begin{equation}
	\begin{split}
		( z_2^{L+1}- \delta z_2)\phi_{z_1}&\approx (z_1^{L+1}- \delta z_1)\phi_{z_2}\ .
	\end{split}
\end{equation}
Moving all the $\delta$ terms to the right-hand side of the equation,
\begin{equation}
	\begin{split}
		z_1^{L+1}\phi_{z_2}-  z_2^{L+1}\phi_{z_1}&\approx \delta (z_1\phi_{z_2}-z_2\phi_{z_1})\ .\\
	\end{split}
\end{equation}
Recalling that $ z_1z_2=t_R/t_L$, we can further simplify the above into 
\begin{equation}
	\begin{split}
		z_1^{L+1}\phi_{z_2}- \frac{1}{z_1^{L+1}}\left( \frac{t_R}{t_L}\right)^{L+1}\phi_{z_1}&\approx\delta (z_1\phi_{z_2}-z_2\phi_{z_1})\ ,\\
	\end{split}
\end{equation}
that is, 
\begin{equation}\label{EqGBZ1}
	z_1^{L+1}\approx \frac{\delta (z_1\phi_{z_2}-z_2\phi_{z_1})}{2\phi_{z_2}}+ \sqrt{\left(\frac{\delta (z_1\phi_{z_2}-z_2\phi_{z_1})}{2\phi_{z_2}}\right)^2 +\left(\frac{t_R}{t_L}\right)^{L+1}\frac{\phi_{z_1}}{\phi_{z_2}}}\ ,
\end{equation}
with the larger root chosen since $|z_2| \leq \sqrt{|t_R/t_L|} \leq |z_1|\leq 1$.

\begin{figure}
	\centering
	\includegraphics[width=0.95\linewidth]{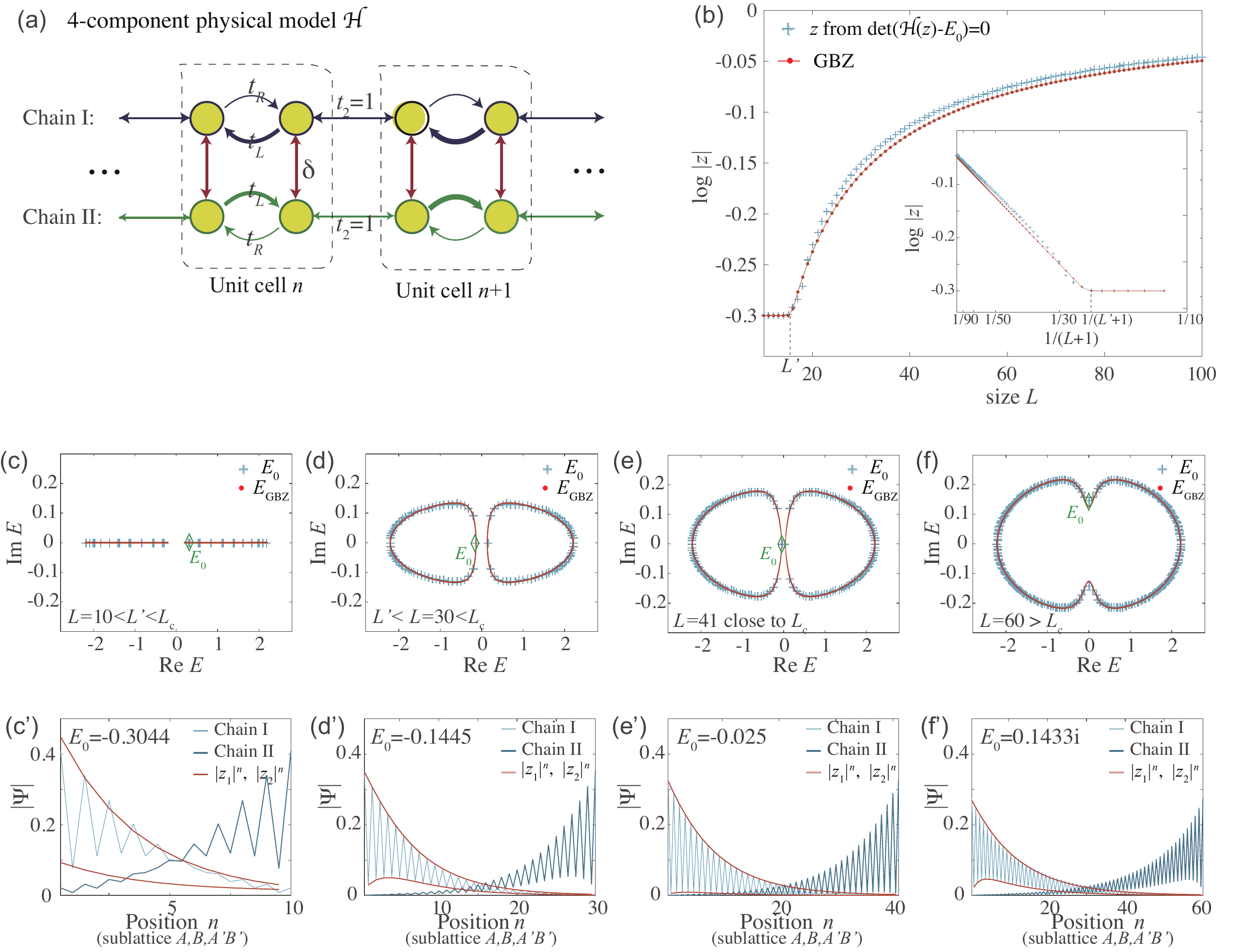}
	\caption{{\bf Scaling-dependent GBZ, spectrum and eigenfunctions of the 4-component model $\mathcal{H}$ (Eq.~\eqref{EqS1}).}
(a) The two chain 4-component model (Eq.~\eqref{EqS1}) incorporates equal but opposite hopping asymmetries $t_L$ and $t_R$ in both chains I and II. Even a small inter-chain coupling $\delta$ can lead to significant scaling dependence when $t_L \neq t_R$. 
(b) Agreement of the scale-dependent skin depth $\log |z|$ obtained via two different ways: numerically, by solving $\det(\mathcal{H}(z)-E_0)=0$ from the numerical eigenspectrum $E_0$ and choosing slowest decaying $z$ solution (blue crosses); analytically, via the scale-dependent generalized Brillouin zone (GBZ) we derived (Eq.~\eqref{EqGBZ4})  (red dots) . $L'$ (Eq.~\eqref{Lp}) represents the system size at which the skin depth (GBZ) transitions from being scale-independent to scale-dependent. 
(c-f, c'-f') Energy spectra at various system sizes $L$ and illustrative  {eigenfunctions} at energies $E_0$ (green diamonds in (c-f)) of smallest absolute value. 
Excellent agreement is observed between the numerical (blue) and analytically-derived GBZ (red) spectra (Eq.~\eqref{EqGBZ4}) and eigenfunctions (Eq.~\eqref{z1z2}, Eq.~\eqref{Eqpsi}). The lower bound for the eigenfunctions (lower red curves) results from the superposition of $z_1$ and $z_2$ solutions in Eq.~\eqref{Eqpsi}, and is generally non-monotonic except at the critical scale $L\approx L_c$ (Eq.~\eqref{Eqcritical1}), where it exhibits a perfectly exponential profile since both $z$ solutions coalesce. Parameters are $t_L = 1.2\text{e}^{0.3}, t_R = 1.2\text{e}^{-0.3}$, and $\delta = 1.6\times 10^{-3}$ for all panels.
	 }
	\label{picS1}
\end{figure}

Below, we highlight 2 extreme but highly relevant cases:
\begin{itemize}
	\item  $ \delta \ll  {t_R^L}/{t_L^L}$, which is the regime where the coupling $\delta$ is so weak  {and imposes a constraint on the system size,}
 that it remains negligible despite the presence of NHSE competition. In this case, only the second term under the square root in Eq.~\eqref{EqGBZ1} survives
	\begin{equation}\begin{split}
			z_1^{L+1}&\approx \frac{\delta (z_1\phi_{z_2}-z_2\phi_{z_1})}{2\phi_{z_2}}+\sqrt{\frac{t_R^{L+1}}{t_L^{L+1}}\frac{\phi_{z_1}}{\phi_{z_2}}} + \frac{1}{2}\left(	\sqrt{\frac{t_R^{L+1}}{t_L^{L+1}}\frac{\phi_{z_1}}{\phi_{z_2}}}\right)^{-1}\left(\frac{\delta (z_1\phi_{z_2}-z_2\phi_{z_1})}{2\phi_{z_2}}\right)^2\\
			&\approx\sqrt{\frac{t_R^{L+1}}{t_L^{L+1}}\frac{\phi_{z_1}}{\phi_{z_2}}}\quad.\end{split}\end{equation}
	As $\sqrt[L+1]{{\phi_{z_1}}/{\phi_{z_2}}}\approx 1$ for moderately large $L$   {($L\geq 10$)}, where $\phi_z=(t_R+z)/E$, we recover
	\begin{equation}
		z_1,z_2\approx \sqrt{\frac{t_R}{t_L}}\text{e}^{\pm ik} \ ,
	\end{equation}
	with $k=n\pi/(L+1)$, $n=1,2,\cdots, L$. This is just the usual GBZ expression for satisfying OBCs $\psi(x=0) = \psi(x=L+1) = 0$ in the uncoupled ($\delta=0$) case.\\
	
	\item  $ \delta \gg {t_R^L}/{t_L^L}$, which defines the strongly coupled regime. Here, the first term under the square root in Eq.~\eqref{EqGBZ1} dominates, such that	
	\begin{equation}\label{eqGBZ2}
		\begin{split}
			z_1^{L+1}&\approx \frac{\delta (z_1\phi_{z_2}-z_2\phi_{z_1})}{\phi_{z_2}}+\left(\frac{\delta (z_1\phi_{z_2}-z_2\phi_{z_1})}{2\phi_{z_2}}\right)^{-1}\frac{t_R^{L+1}}{t_L^{L+1}}\frac{\phi_{z_1}}{\phi_{z_2}}\\	
			&\approx  \frac{\delta (z_1\phi_{z_2}-z_2\phi_{z_1})}{\phi_{z_2}} \ .
	\end{split}\end{equation}

In principle, the above expression can be exactly solved to yield the GBZ (i.e. $z_1$), which we see must depend on the system size $L$. However, because $\phi_{z_1},\phi_{z_2}$ depend on $E$, which further depends on the system parameters in a complicated manner, it is useful to perform some approximations such that the right-hand side does not depend explicitly on $E$. One convenient series of approximations is 
	\begin{equation}\label{eqGBZ3}
		\begin{split}
			& z_1=\left(\delta\times \frac{(z_1\phi_{z_2}-z_2\phi_{z_1})}{\phi_{z_2}}\right)^{\frac{1}{L+1}}\text{e}^{ik} \\
			&\xrightarrow{\sqrt[L+1]{{\phi_{z_2}}/{\phi_{z_1}}}\approx 1} \left(\delta\times \frac{(z_1\phi_{z_2}-z_2\phi_{z_1})}{\phi_{z_1}}\right)^{\frac{1}{L+1}}\text{e}^{ik} \\ 
			&\xrightarrow{\phi_z=(t_R+z)/E} \left(\delta \times \frac{t_R( t_L z_1- t_R/z_1)}{t_L(t_R+z_1)}\right)^{\frac{1}{L+1}}\text{e}^{ik} \ .
	\end{split}\end{equation}
While this expression technically needs to be solved self-consistently for $z_1$, noting that the right-hand side depends very slowly due to the $1/(L+1)$ exponent, we can simply approximate $z_1$ in it with suitable constants i.e. $z_1\approx -1$ such as to obtain
	\begin{equation}
		z_1\approx\left( \delta \times\frac{t_R( t_R- t_L)}{t_L(t_R-1)}\right)^{\frac{1}{L+1}}\text{e}^{ik}  \ ,
	\end{equation}
	with $k=2n\pi/(L+1)$, $n=1,2,\cdots, L$. 
	
	This approximate expression for the GBZ is dependent on the system size $L$ in the form of the exponent $1/(L+1)$, which gives weak dependence on the inter-chain coupling $\delta$ as well as a constant factor containing the other system parameters. While it is not the only possible approximation, it is relatively compact and agrees well with the numerical GBZ values  (Fig.~\ref{picS1}b) obtained by substituting the exact OBC spectrum $E_0$ of our 4-component physical system into $\text{det}(\mathcal{H}(z)-E_0\mathbb{I})=0$, and solving for $z$. Also, it predicts a spectrum $E_\text{GBZ}=\pm \sqrt{(t_L+1/z_1)(t_R+z_1)}$ that agrees very well with that numerically-obtained exact OBC spectrum $E_0$, as shown in Fig.~\ref{picS1}c-f. Regardless of the exact choice of approximation, the $\delta^{1/(L+1)}$ dependence holds universally, and suggests a slow but $\delta$-dependent convergence of the GBZ onto the unit circle in the $L\rightarrow \infty$ limit. In Fig.~\ref{picS1}c'-f', the $|z_{1}|^{x},|z_{2}|^{x}$ spatial decay profiles predicted by the GBZ (brown) near $E_0=0$ also exhibit excellent agreement with the upper and lower envelope limits of the numerical eigenstate $\Psi$ closest to $E_0=0$ in Fig.~\ref{picS1}c-f.
\end{itemize}

   {For different values of $\delta$ and system sizes $L$, we assume that the GBZ changes slowly and continuously.}
In all, our approximate size-dependent GBZ solution for our coupled HN model system is given by
\begin{equation}\label{EqGBZ4}
	\begin{split}
		&\  z =\exp(iK)=\left \{\begin{array}{ll}  
			\sqrt{t_R/t_L} \text{e}^{ik}, \qquad  &\qquad L \leq L' \\
			\\
			\text{e}^{-\frac{\alpha}{L+1}} \text{e}^{ik}, &\qquad L \geq L'\\
		\end{array}\right.\ ,\qquad \alpha = -\log \left|\delta \times \displaystyle\frac{ t_R( t_R- t_L)}{t_L(t_R-1)}\right|\ ,\\
	\end{split}
\end{equation} 
where 
\begin{equation}
L'=-\alpha/\log\sqrt{t_R/t_L}-1\ ,
\label{Lp}
\end{equation}
is the transition length in which the energy spectrum or skin depth $-\log|z|$ switches from being size-independent to size-dependent (Fig.~\ref{picS1}b). The value of $L'$ is determined by setting both expressions for the GBZ to be equal i.e. $|z_1|=\sqrt{t_R/t_L} = \exp\left(-{\alpha}/({L'+1})\right)$. The very good agreement between our analytic GBZ solution (Eq.~\eqref{EqGBZ4}) and exact numerics is presented in Fig.~\ref{picS1} for the skin depth $-\log|z|$, spectrum $E$ and eigenfunction decay.

In a nutshell, we have shown that as the system size $L$ increases, the GBZ and spectrum always switches from being size-independent to size-dependent as $L$ crosses $L'$. This is given by our approximate GBZ expression Eq.~\eqref{EqGBZ4} which nevertheless exhibits excellent agreement with numerical results from exact diagonalization. Our approach was based on neglecting the effect of weak coupling terms in the bulk equations but retaining their influence in enforcing the boundary conditions, which allowed for analytic headway in writing down the energy expression from the characteristic equation of a single chain $\det(H(z)-E)=0$. This is justified because while extremely weak couplings should not significantly affect the bulk, they would still nontrivially interplay with the boundary conditions due to the high sensitivity of non-Hermitian systems to spatial inhomogeneities. 
To conclude, with nonzero coupling between chains with antagonistic NHSE, the GBZ is found to be no longer given by the usual $|z_1|=|z_2|$ condition, but instead takes on a $L$-dependent behavior that crucially involves the coupling strength $\delta$.


\subsection{ Effective size-dependent two-component model }\label{suppmat1c}


The key purpose of any GBZ construction is to ``remove'' the NHSE, such that the Hamiltonian evaluated on the GBZ (i.e. ``surrogate Hamiltonian''~\cite{lee2020unraveling}) can already accurately reproduce the OBC spectrum without actually diagonalizing it under OBCs. In our case, the size-dependent GBZ obtained in the previous subsection has already encapsulated the effects of the coupling and the antagonistic NHSE.

As such, working in our effect size-dependent GBZ $z$, we can accurately study $\mathcal{H}$ (Eq.~\eqref{EqS1}) as two uncoupled copies of 2-component chains $H$ i.e. $\mathcal{H} = H(K) \oplus H^{T}(-K)$, where $K=-i\log z$. The effects of the coupling and NHSE are implicit in the imaginary part of $K$. From Eq.~\eqref{EqS2},
\begin{equation}\label{Eqeffectivemodel} 
H(K) = \begin{pmatrix} 
		0 & t_L + \text{e}^{-iK}\\
		t_R+ \text{e}^{iK}& 0
	\end{pmatrix}, 
\end{equation}
even though our approach would work well for generic $H(k)$ of similar levels of complexity. To recall, 
\begin{equation}\label{Eqeffectivemodel1} 
K=-i\log z=k+i\frac{\alpha}{L+1}\ ,
\end{equation}
from Eq.~\eqref{EqGBZ4}, where the real part $ k=\text{Re} K=2m\pi/(L+1)$ with $m=1,2,...,L$, and the imaginary part $\text{Im} K=\displaystyle\frac{\alpha}{L+1}$ is controlled by 
\begin{equation}\label{Eqeffectivemodel2} 
\alpha =   \left \{\begin{array}{ll}-\log \left|\delta \times \displaystyle\frac{t_R(t_R-t_L)}{t_L(t_R-1)}\right|, &\qquad L\geq L'\\
	-(L+1)\log \sqrt{t_R/t_L} \ ,&\qquad L\leq L'
\end{array}\right. \ ,
\end{equation}
from Eqs.~\eqref{EqGBZ4} and \eqref{Lp}. From direct diagonalization of $H(K)$, the eigenenergies are given by
\begin{equation}
	\begin{split}
		&\mathcal{E}=\pm \sqrt{(t_L+\exp(-iK))(t_R+\exp(iK))}\ ,\\
	\end{split}
	\label{2bandenergy}
\end{equation} 
which has a similar form as Eq.~\eqref{Ez}; but when evaluated on the GBZ (Eq.~\eqref{EqGBZ4}), it should reproduce the OBC spectrum of the entire original 4-component model $\mathcal{H}$ (Eq.~\eqref{EqS1}). Note that we have used the notation $\mathcal{E}$ to refer to the eigenenergies predicted by the GBZ, which are supposed to closely approximate the true eigenenergies $E$.

Importantly, because of the size dependence of the GBZ, $H(K)$ can possess exceptional points (EPs) at special sizes $L=L_c$ where the corresponding $K=K_c$ gives rise to $t_R+\exp(iK_c)=0$ and $t_L+\exp(-iK_c)\neq 0$. When that occurs, $H(K)$ is of the Jordan form $\sigma_+$ and is not of full rank (defective). We will soon investigate how this scaling-induced EP lead to unconventional entanglement behavior.
Substituting $K_c=k_c +i\alpha/(L_c+1)$,
\begin{equation}
	\begin{split}\label{Eqcritical1}
		&t_R+\exp(iK_c)=t_R+\exp(-\alpha/(L_c+1)+ik_c)=0 \\
		&\Rightarrow k_c=\pi ,\qquad  {\alpha}/{(L_c+1)}=-\log t_R\\
		& \Rightarrow L_c=-\frac{\alpha}{\log t_R}-1\ ,
	\end{split}
\end{equation} 
with $t_{R}, t_L>0$ and  $\alpha = -\log \left|\delta \times \displaystyle\frac{t_R(t_R-t_L)}{t_L(t_R-1)}\right| $. 
Here, $L_c$ represents the critical length in which an EP appears, and should not be confused with $L'=-\alpha/\log\sqrt{t_R/t_L}-1$ (Eq.~\eqref{Lp}), which is the length above which the GBZ becomes size dependent. Evidently, such a scaling-induced EP exists only if $L_c>L'$.

\begin{figure}[H]
	\centering
	\includegraphics[width=.75\linewidth]{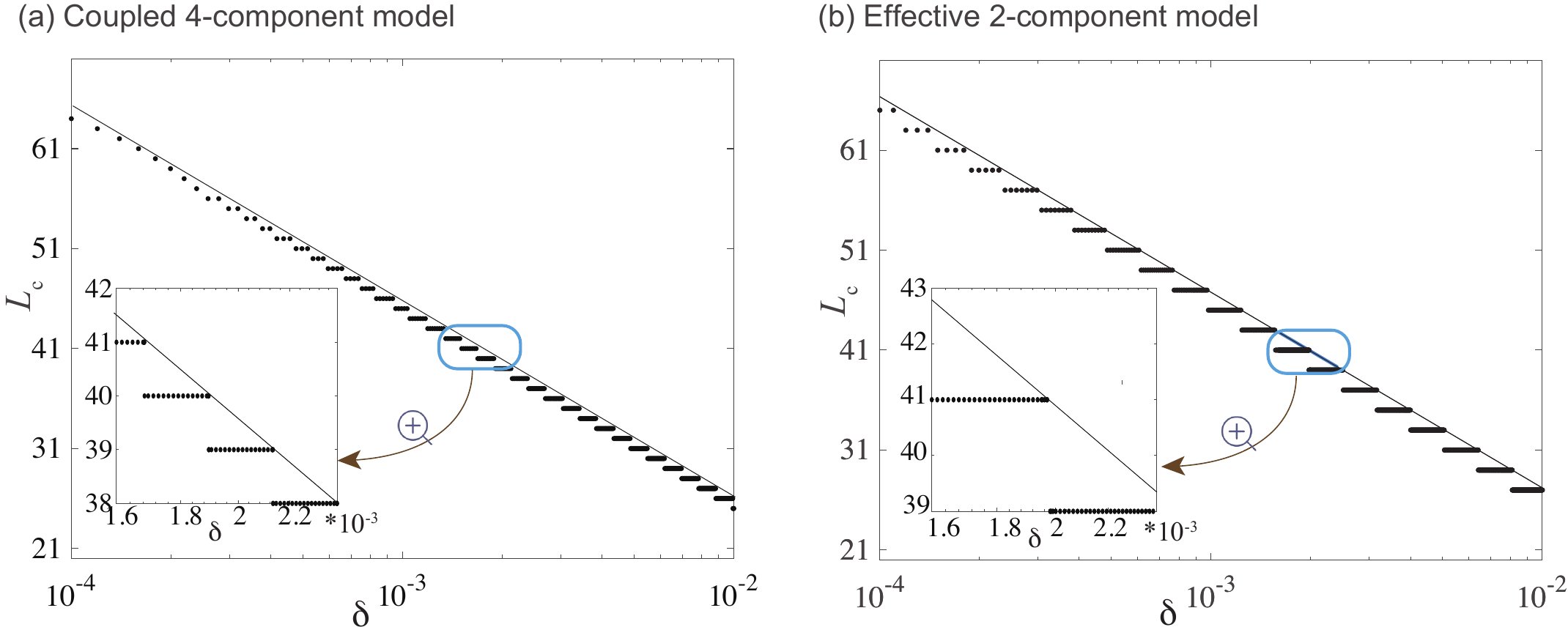}	
	\caption{{\bf The critical system size $L_c$ dependence on $\delta$ for the coupled 4-component model (Eq.~\eqref{EqS1}) and its effective two-component approximation (Eq.~\eqref{Eqeffectivemodel}).}
The spectral gap of our model closes at an EP	at the critical system size $L_c$, which is analytically predicted to depend logarithmically on the small inter-chain coupling parameter $\delta$. Excellent agreement is observed between the numerically determined integer $L_c$ (black dots) and its analytical prediction (Eq.~\eqref{Eqcritical1}). Since $\delta$ is built into the GBZ for the two-component, single-chain approximation (Eq.~\eqref{Eqeffectivemodel}), its excellent numerical agreement is testimony to the validity of our scale-dependent GBZ approach. 
}
	\label{picS2}
\end{figure}

Even though $L_c$ was derived from the approximately-obtained GBZ, it already predicts the scaling-induced EP transition to excellent accuracy, when compared to the critical $L_c$ obtained from exact diagonalization spectra (which must be an integer), as shown in Fig.~\ref{picS2}. 

\subsubsection{ Critical behavior near the scaling-induced EP}\label{EEP}
To prepare for the study of the entanglement properties near the scaling-induced EPs described above, we describe how the Hamiltonian behaves in their neighborhood. We first expand the size-dependent quasi-momentum $K$, both its real and imaginary parts, as $\Delta_K=K-K_c$ where $K=k+i\alpha/(L+1)$ and $K_c=\pi+ i\alpha/(L_c+1)$ (See Eqs.~\eqref{EqGBZ4} and \eqref{Eqcritical1}), such that $\Delta_K=\text{Re}\Delta_K+i\text{Im}\Delta_K$ is given by
\begin{align}
\text{Re}\Delta_K&= k-\pi\notag\ ,\\
\text{Im}\Delta_K&= \alpha\left( \frac{1}{L+1}-\frac{1}{L_c+1}\right)\ .
\end{align}
With these, the Hamiltonian (Eq.~\eqref{Eqeffectivemodel}) can be expanded near $K=K_c$ as
\begin{equation}
	\begin{split}
		H(K) &\xrightarrow{\Delta_K=K- K_c } \begin{pmatrix} 
			0 & t_L + \text{e}^{-iK_c}\text{e}^{-i\Delta_K}\\
			t_R+ \text{e}^{iK_c}\text{e}^{i\Delta_K}& 0
		\end{pmatrix}\\
		&\xrightarrow[t_R + \exp(iK_c) = 0]{\text{e}^{i\Delta_K}\approx 1+i\Delta_K} \begin{pmatrix} 
			0 & t_L - 1/t_R \\
			-it_R\Delta_K& 0
		\end{pmatrix}\ ,\\
	\end{split}
\end{equation}
   {due to $ t_L - 1/t_R\neq 0$}.
This yields the eigenenergies of $H(K_c+\Delta_K)$ to be $\pm \mathcal{E}=\pm \sqrt{i( 1-t_L t_R)\Delta_K}$.

\subsubsection{ Critical properties of the occupied band projector $P_K$ }
The geometric defectiveness of an EP is reflected in the divergences of the occupied band projector $P$  {for states with Re$E\leq 0$}, since the latter will not be well-defined with the occupied and valence bands merged into one. For a 2-component model, $P=\sum_k P_k|k\rangle\langle k|$ is defined (in the biorthogonal basis) as
\begin{equation}
\label{EqPk0}
	\begin{split} 
		P_K =\frac{1}{2} \left(\mathbb{I}-\frac{H(K)}{\mathcal{E}(K)}\right)\ ,\end{split} 
\end{equation} 
where $\mathcal{E}(K)$ is the eigenvalue of the Hamiltonian $H(K)$ (Eq.~\eqref{Eqeffectivemodel}). Note that we have analytically continued $P_k\rightarrow P_K$ into the GBZ. 
Ordinarily, away from gap closure points, $P(K)$ projects onto a well-defined occupied band since $\mathcal{E}(K)\neq 0$. But at an EP ($K=K_c$)where $\Delta_K=0$, $ \mathcal{E}(K)= \sqrt{i( 1-t_L t_R)\Delta_K}$ vanishes, rendering $P_K$ singular.

\begin{figure}[H]
	\centering
	\includegraphics[width=.8\linewidth]{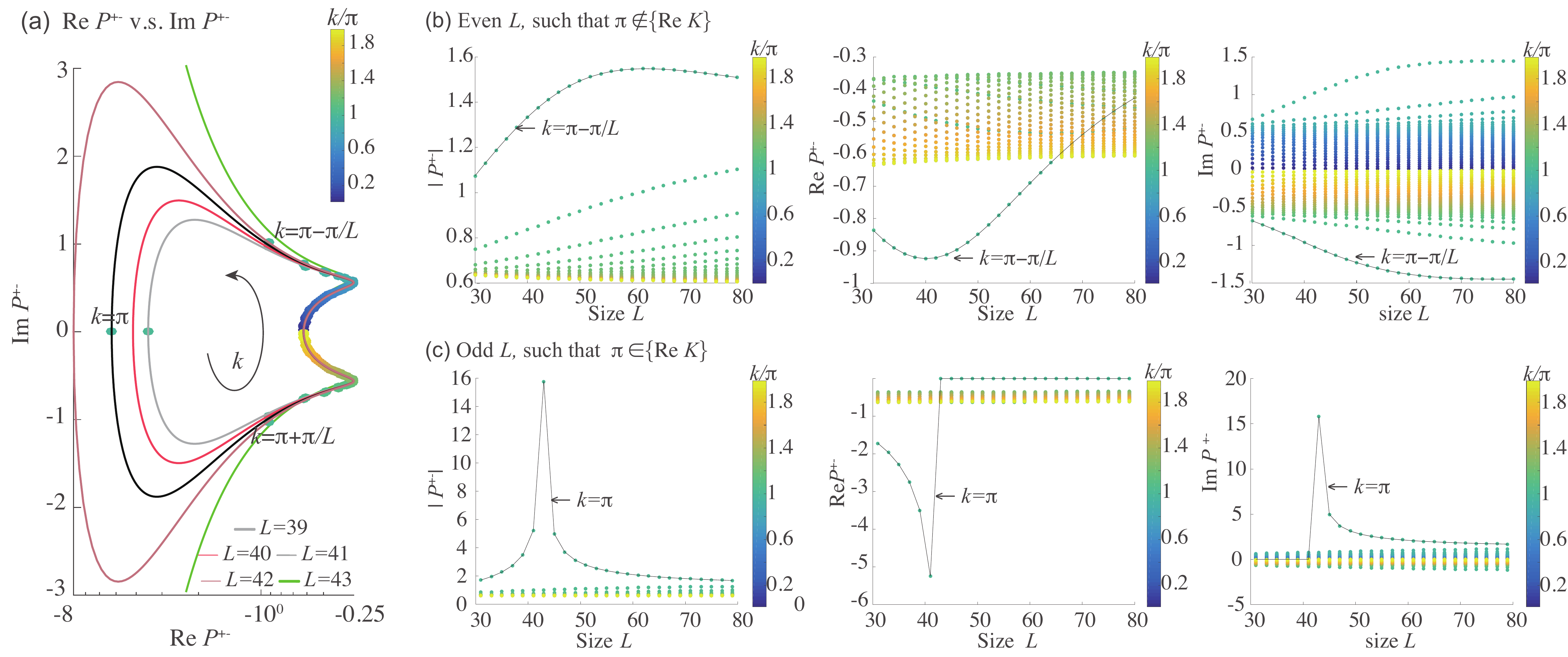}
\caption{{\bf Potentially divergent matrix element $P_K^{+-}$ (Eq.~\eqref{EqPk}) of our truncated projection operator $P_K$ (Eq.\eqref{EqPk0}) 
in even vs. odd $L$ systems.}
	(a) Complex values of the off-diagonal element $P_{K(k)}^{+-}$, which traces a different loop for each system size $L$ as $k$ cycles over a period $[0,2\pi)$ with discrete points $ (2m-1)\pi/L$, $m=1,2,...,L$ (drawn as solid dots). The $k=\pi$ point exist only for odd $L$, and causes $P_{K(k)}^{+-}$ to diverge to infinity as $L\rightarrow L_c=-\frac{\alpha}{\log t_R}-1=42.79$ (Eq.~\eqref{Eqcritical1}), $\alpha= -\log \left|\delta \times \displaystyle\frac{ t_R( t_R- t_L)}{t_L(t_R-1)}\right|$.
	(b, c) The values of $|P_{K(k)}^{+-}|$, Re$(P_{K(k)}^{+-})$ and Im$(P_{K(k)}^{+-})$ as a function of system size $L$, plotted as separate branches for each $k$ value for (b) even and (c) odd $L$. While there exists for even $L$ a $k=\pi-\pi/(L+1)$ branch that departs from the other branches, for odd $L$, the $k=\pi$ branch exhibits a sharp kink for $L=43\approx L_c = 42.79$.
	Parameters are $t_L = 1.2\text{e}^{0.3},t_R = 1.2\text{e}^{-0.3},\delta = 1.6\times 10^{-3}$.}
		\label{picS3}
\end{figure}

   {Closest} to an EP, where  {both the real and imaginary parts of} $\Delta_K$ is very small, 
\begin{equation}\label{EqPk}
	\begin{split} 
		P_K =\frac{1}{2} \begin{pmatrix}1 & P^{+-}_K\\P^{-+}_K & 1\end{pmatrix}\ ,\quad  {P^{+-}_K}=\frac{1}{{P^{-+}_K}}\ , \end{split} 
\end{equation} 
where
\begin{equation}\label{EqP+}
	P^{+-}_K\approx -\frac{\sqrt{t_L/t_R -1/t^2_R}}{2\sqrt{-i\Delta_K}}\quad ,
\end{equation} 
with $\Delta_K=K-K_c=k - \pi + i(\frac{\alpha}{L+1}-\frac{\alpha}{L_c+1})$. 
In a finite system, the the quasi-momentum $k=\text{Re}K$ points take the values $2\pi/(L+1),4\pi/(L+1),...$, such that Re$\Delta_K\sim \pi/L$ for even-sized systems, and Re$\Delta_K=0$ for odd-sized systems.
However, the imaginary momentum deviation Im$\Delta_K$ is approximately $\frac{\alpha}{L+1}-\frac{\alpha}{L_c+1}=\frac{\alpha(L_c-L)}{(L+1)(L_c+1)}\sim \alpha(L_c-L)/L^2\approx \alpha(L_c-L)/L_c^2$ near $K_c$, which is much smaller than $1/L$. Since $P^{+-}_K\propto \Delta_K^{-1/2}$, the strength of the singularity depends on both the real and the imaginary parts of $\Delta K$, as illustrated in the complex $P_+$ plot in Fig.~\ref{picS3}a ($L=43$ is the most divergent). To elaborate on the qualitatively different cases of even and odd $L$: 


\begin{itemize}
	\item Even $L$, such that min$|\text{Re}\Delta_K|\neq 0$:\\
	\indent Here, the closest momentum point passes approximately within $\pi/L$ of Re$K_c=\pi$, such that 
$|\Delta_K|=\left|\frac{\pi}{L}+i\left(\frac{\alpha}{L_c+1}-\frac{\alpha}{L+1}\right)\right|\approx\frac{\pi}{L}$ for modest system sizes of  $L,\ L_c\ \sim\ 10-10^2$.  The scaling behavior is dominated by $|\Delta_K|\sim $Re$\Delta_K$-- masking the effect of the imaginary part. As shown in Fig.~\ref{picS3}b, the most divergent contribution to $P^{+-}_K\sim \Delta_K^{-1/2}\sim L^{1/2}$ arises from $k=\pi-\pi/L$, but because $|\Delta_K|$ diverges with $L$ and not $L_c$, it does not really capture the value of $L_c$.

	\item  Odd $L$, such that min$|\text{Re}\Delta_K|= 0$:\\
	\indent Here, $k=\pi$ is visited in the GBZ, and Re$\Delta_K=0$ while $\Delta_K=i\left(\frac{\alpha}{L+1}-\frac{\alpha}{L_c+1}\right)\sim \alpha(L_c-L)/L^2_c$. We hence have $|P^{+-}_K|\sim \Delta_K^{-1/2}\sim L_c/\sqrt{\alpha|L_c-L|}$, which indeed shows up as a divergence at $L\approx L_c$ in the $P^{+-}$ plots in Fig.~\ref{picS3}c. 
\end{itemize}

The fundamental difference between our scaling-induced EPs and usual critical points is that the divergence in the projector does not stem from Re$\Delta_K$, which typically scales like $L^{-1}$, but instead arises from Im$\Delta_K$, which diverges at $L\approx L_c$.


\section{ Free-fermion entanglement entropy dip due to scaling-induced EPs}\label{suppmat2}


In the previous section, we have derived the system-size (scaling) dependent GBZ that captures the effects of coupling two chains with antagonistic NHSE, and used that to accurately approximate our original 4-component system with an effective 2-component (single-chain) Hamiltonian. The key conclusion was that, for odd system sizes $L$, the truncated occupied band projector $\bar P$ is expected to diverge at a special $L=L_c$ (in practice, the computed $L_c$ is usually not an integer, and $\bar P$ becomes very large though finite.) 

In this section, we show how this divergence lead to an anomalous dip in the entanglement entropy (EE) scaling of free fermions, and how that can be analytically estimated and characterized. We note that this EE scaling is atypical because, in usual cases where the non-Hermiticity is isotropic, 
quantum correlations will spread evenly throughout the system, mirroring the distribution seen in Hermitian quantum systems. Even in prototypical NHSE systems where the Hermiticity can be gauged away through a basis transform. the hopping asymmetry does not lead to nontrivial modifications to the EE. In our case however, the EE is shown not to conform to standard area or volume laws but instead exhibit an unique entanglement dip.


For  {free fermions}, the entanglement spectrum can be obtained by truncating the occupied band projector $P$ in real-space. In 1D, we define a real-space partition $[x_L, x_R]$, such that truncated band projector $\bar P$ can be obtained from $P$ via
\begin{equation}\begin{split}\label{2bandP3}
		& \bar P=  \bar{\mathcal{R}}_{[x_L, x_R] } P  \bar{\mathcal{R}}_{[x_L, x_R] } \ ,\ \\
		& \bar{\mathcal{R}}_{[x_L, x_R] }=\sum^{L}_{x{\notin}[x_L, x_R] }|x\rangle\langle x|\otimes \mathbb{I}\ ,\\
		&P= \sum_{n,(\text{Re}E_n)<0}|\psi_n^R\rangle \langle \psi_n^L|\ ,
\end{split}\end{equation}
where $\bar{\mathcal{R}}_{[x_L, x_R] }$ is the projector onto sites outside of $[x_L, x_R] $ (note that the index $n$ was expressed as indices $(k,\mu)$ in the main text, such as to emphasize the momentum $k$). For two-component models, the above general expression for $P$ reduces to Eq.~\eqref{EqPk0}. 
The eigenvalues of the projection operators can be interpreted as occupancy probabilities. For $P$, they are limited to either 1 and 0. However, when the system is divided in to the $[x_L, x_R]$ and its complement, the truncated occupied matrix $\bar P$ reveals 
 {the entanglement} between these two regions, resulting in occupancy probabilities (eigenvalues) $p$ away from $1$ and $0$.  
The free-fermion entanglement entropy (EE) $S$ quantifies this entanglement, and is given by
	\begin{eqnarray}
	\label{SEE}
		S(\bar P) &=& -\text{Tr}(\bar P \log \bar P +(\mathbb{I}-\bar P)\log (\mathbb{I}-\bar P)) \notag\\
		&=&-\sum_p (p \log p +(1-p)\log (1-p)) \ .
	\end{eqnarray}

\subsection{Entanglement entropy scaling and dip in the effective 2-component model Eq.~\eqref{Eqeffectivemodel} }
\label{suppmat2a}

To investigate the EE behavior, we first compute $\bar P$ in its real-space basis, such that $\bar{\mathcal{R}}_{[x_L, x_R] }$ can be implemented as the truncation to a submatrix. Employing the Fourier Transform $|x,\alpha\rangle =\frac{1}{\sqrt{L}}\sum_k\text{e}^{-ikx}|k,\alpha\rangle$, we can express its real-space matrix elements as 
\begin{equation}\label{eq2cor}
	\begin{split}
		\langle c^{\dagger}_{x_1,\alpha}c_{x_2,\beta}\rangle=	\langle {x_1,\alpha}| \bar P|x_2,\beta\rangle=\left\{\begin{array}{cc}\displaystyle\frac{1}{L}\sum\limits_k P^{\alpha\beta}_K\text{e}^{ik(x_1-x_2)}, &\qquad x_1,x_2\notin[x_L, x_R] \\
			0 &\qquad x_1,x_2\in[x_L, x_R]
		\end{array}\right.\  ,
\end{split}\end{equation} 
where $\alpha,\beta=\pm$ represent the sublattices $A,B$, and $K=k+i\alpha/(L+1)$. Note that even though $P_K$ is evaluated in the GBZ $K$, the fourier transform is still between the real-space lattice and the usual BZ $k$.

\begin{figure}[H]
	\centering
	\includegraphics[width=0.75\linewidth]{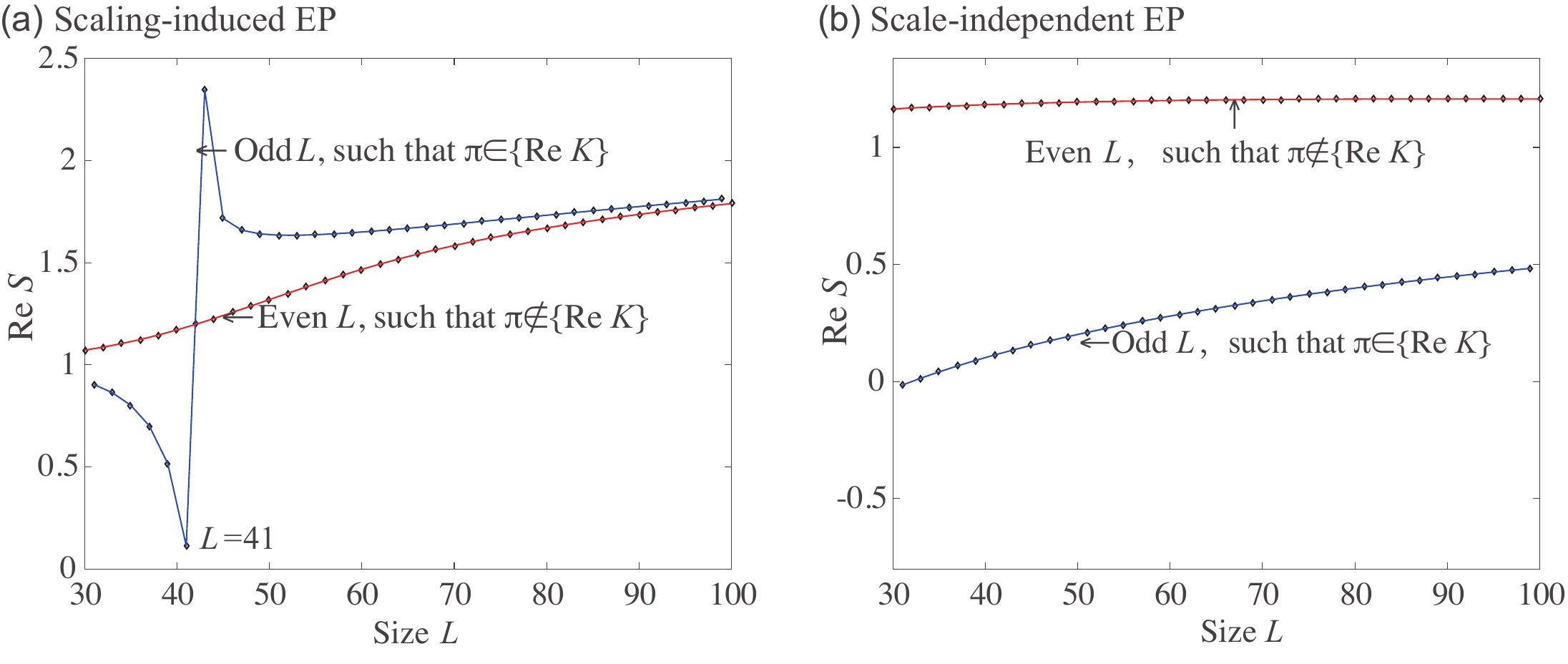}
	\caption{
\label{picS4} 
{\bf Comparison of entanglement entropy scaling behavior with and without scaling-induced exceptional criticality.} 
(a) For our 2-component effective model (Eq.~\eqref{Eqeffectivemodel1}) with scale-dependent GBZ given by Im$K = \frac{\alpha}{L+1}$, the entanglement entropy experiences a pronounced dip at $L=41\approx L_c$. This only occurs for odd $L$, where the Re$K=\pi$ point is sampled. 
(b) Had the same model Eq.~\eqref{Eqeffectivemodel1} been a physical model with fixed GBZ, rather than an effective model of a $L$-dependent GBZ, there would not have been any entanglement dip. We have fixed Im$K= -\log \left|\delta \times \displaystyle\frac{ t_R( t_R- t_L)}{t_L(t_R-1)}\right|/(L_c+1) =-0.123$, with parameters $t_L = 1.2\text{e}^{0.3},t_R = 1.2\text{e}^{-0.3}, \delta=1.6\times 10^{-3}$. 
	}
\end{figure}

Because of the peculiar scaling properties of our GBZ of our effective 2-band model Eq.~\eqref{Eqeffectivemodel}, $\bar P$ behaves very differently for odd and even $L$ (See Fig.~\ref{picS4}). For odd $L$, we have shown that the discretized momenta in the GBZ contain one point where Re$K =\pi$, i.e. min$|\text{Re}\Delta_K|$ from the EP is zero, where $\Delta_K=i\left(\frac{\alpha}{L+1}-\frac{\alpha}{L_c+1}\right)$. In this case, $P^{+-}_K$ (Eq.~\eqref{EqP+}) is divergent as $|L-L_c|^{-1/2}$ near the EP. However, for even $L$, the divergence is only asymptotic with $L^{1/2}$, as explained below Eq.~\eqref{EqP+}. In the following, we elaborate on the above
\begin{equation}\begin{split}\label{pp}
		\langle {x_1,+}| \bar P|x_2,-\rangle 
		&\approx \displaystyle\frac{ {1-(-1)^L}}{2L} P^{+-}_{K=\pi+i\frac{\alpha}{L+1}}\text{e}^{i\pi(x_1-x_2)} +\frac{1}{\pi}\int ^{2\pi}_{\pi+\pi/L}P^{+-}_{K=k+i\frac{\alpha}{L+1}} dk.
\end{split}\end{equation}
Here we have split the integral contributions into two parts: the first term is from $\text{Re}\Delta_K=\pi$ and exists only when $L$ is odd, while the second term contains all other momentum point contributions. From the arguments below Eq.~\eqref{EqP+}, the first term behaves like
 {
\begin{equation}\begin{split}\label{pp1}
		 \displaystyle\frac{ {1-(-1)^L}}{2L} P^{+-}_{K=\pi+i\frac{\alpha}{L+1}}\text{e}^{i\pi(x_1-x_2)}&\approx  -\frac{ {1-(-1)^L}}{2L}\frac{\sqrt{t_L/t_R -1/t^2_R}}{2\sqrt{-i\Delta_K}}\text{e}^{i\pi(x_1-x_2)}\\
		&\xrightarrow[L\rightarrow L_c]{\Delta_K=K-K_c,\,\text{Re}\Delta_K=0\quad}\quad\frac{ {1-(-1)^L}}{2L}\frac{\sqrt{t_Lt_R -1}}{2t_R}\left(\frac{(L+1)(L_c+1)}{\alpha (L_c-L)}\right)^{1/2}\text{e}^{i\pi(x_1-x_2)}\\
		&\xrightarrow{L\approx \sqrt{(L_c+1)(L+1)}}\qquad\quad( {1-(-1)^L})\frac{\sqrt{t_Lt_R -1}}{4\sqrt{\alpha}\cdot t_R} (L_c-L)^{-1/2}\text{e}^{i\pi(x_1-x_2)}\ .\\
\end{split}\end{equation}}
The second term behaves like 
\begin{equation}\begin{split}\label{pp2}
\frac{1}{\pi}\int ^{2\pi}_{\pi+\pi/L}P^{+-}_{k+i\frac{\alpha}{L+1}} dk
&\approx \frac{1}{\pi}\int^{\pi}_{\frac{\pi}{L}}\frac{\sqrt{t_L/t_R -1/t^2_R}}{2\sqrt{-i\Delta_K}}\text{e}^{i(\pi+\text{Re}\Delta_K)(x_1-x_2)}\text{d} \text{Re}\Delta_K\ \propto L^{-1/2}\ .
				\end{split}\end{equation}
In the above, we have not focused on the spatial $x_1-x_2$ dependence, since we are primarily concerned about the $L$-scaling behavior. Diagonalizing $\bar P$, it is numerically verified (See Fig.~\ref{picS4}a) that due to our scaling-induced EP, odd and even-sized systems behave qualitatively differently. Near the critical size, the EE (computed from the eigenvalues $p$ of $\bar P$ via Eq.~\eqref{SEE}) for odd $L$ exhibits a divergence which we call an \emph{entanglement dip}, while in even-$L$ systems, it changes continuously with size.  

However, this entanglement dip disappears if the system does not possess such a scaling-dependent GBZ. If we were to instead consider Eq.~\eqref{Eqeffectivemodel} to be the physical (not effective) model for a non-antagonistic NHSE chain which exhibits a scaling-independent EP point, such that \(K = k + i\frac{\alpha}{L_0 + 1}\) with fixed $L_0$ (instead of $L$),  its EE would no longer show abrupt changes with size, despite still exhibiting odd/even system size effects (See Fig.~\ref{picS4}b).\\

\begin{figure}[b]
	\centering
	\includegraphics[width=0.95\linewidth]{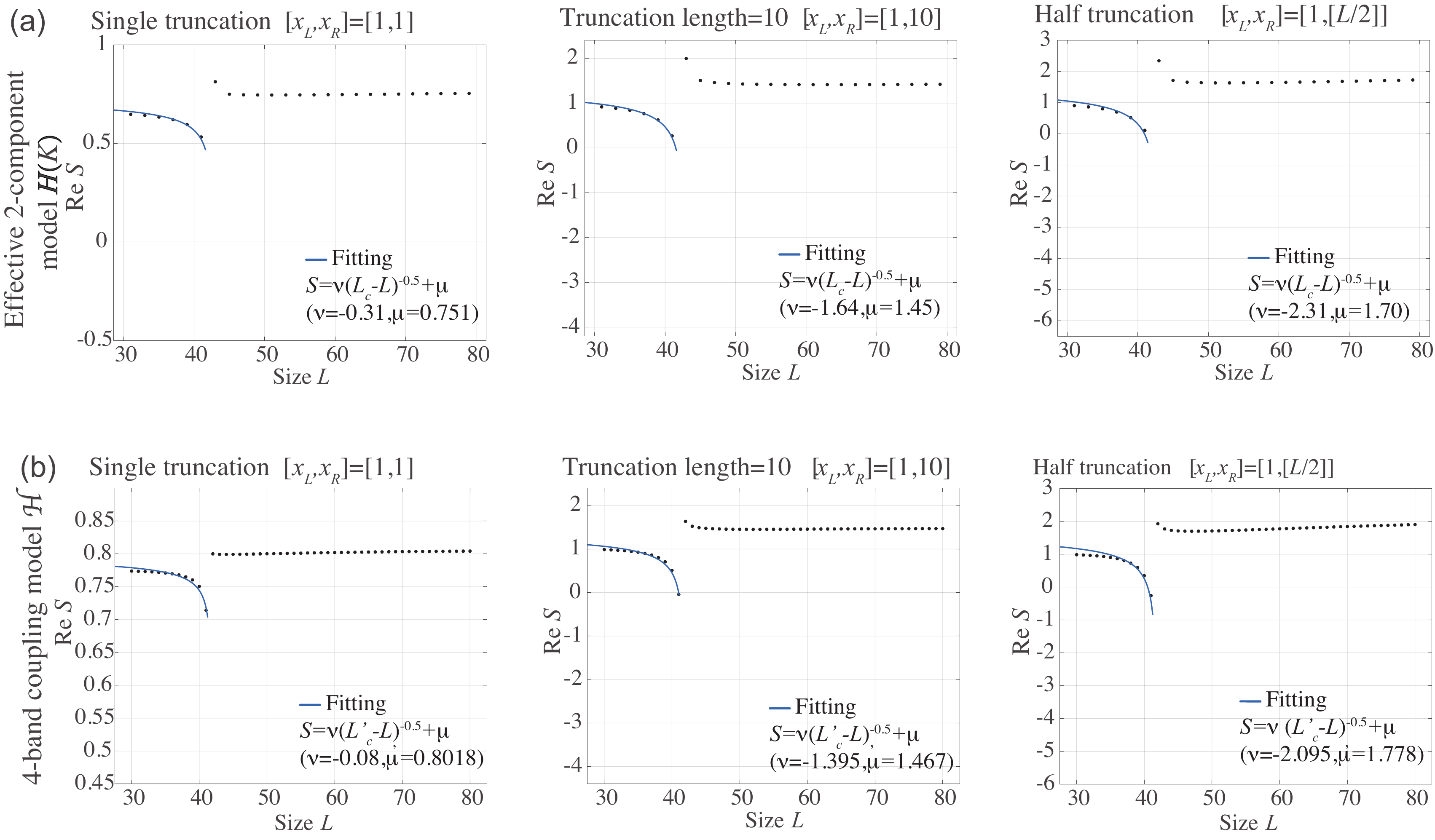}
	\caption{	\label{picS5}
		{\bf Entanglement dip with different entanglement cuts.}
		For different truncated regions $[x_L,x_R]$ implemented by the operator ${\bar{\mathcal{R}}}_{[x_L, x_R] }=\sum^{L}_{x\notin[x_L, x_R] }|x\rangle\langle x|\otimes \mathbb{I}$, similar behavior is observed in the entanglement entropy (EE) of (a) our effective 2-component model $H(K)$ Eq.~\eqref{Eqeffectivemodel} and
		(b) its parent 4-component coupled chain
		model $\mathcal H$ Eq.~\eqref{EqS1}. For all cases, a prominent entanglement dip occurs, with approximate asymptotic $(L_c-L)^{-0.5}$ behavior. 
		Parameters are $t_L = 1.2\text{e}^{0.3},t_R = 1.2\text{e}^{-0.3},\delta=1.6\times 10^{-3}$, corresponding to $L_c=42.79$ for (a) and $L'_c=41.8$ for (b). 
		}
\end{figure}



\subsection{ Effect of entanglement truncation interval in the effective 2-component model }\label{suppmat2b}

The entanglement dip at $L\approx L_c$ occurs universally, regardless of the actual entanglement cut region $[x_L,x_R]$, since it is not specifically dependent on the real-space profile of the matrix elements of $\bar P$. In Fig.~\ref{picS4}a for our 2-component model Eq.~\eqref{Eqeffectivemodel}, the odd-sized EE is shown to consistently exhibit $(L-L_c)^{-0.5}$ divergence regardless of the truncation region. 

This entanglement dip remains qualitatively similar (as it should be) in the EE of its parent 4-component coupled system (Eq.~\eqref{EqS1}), even though the GBZ takes on a more intricate structure and the real part of the EP momentum is not rigorously fixed at $\pi$. Indeed, for this 4-component coupled system, the entanglement dip is observed in both odd and even-sized systems (See Fig.~\ref{picS4}b), implying that it is not fundamentally based on the positions of the $\text{Re}K$ momentum points. What is  fundamental is the fact that the EP lies on a GBZ whose imaginary part of momentum deviation $\Delta_K$ gives rise to a $(L-L_c)^{-0.5}$ divergence in $P$.

\subsubsection{Single unit-cell truncation }

Since the entanglement dip is largely independent of the truncation region, we can make further analytic progress by focusing on \emph{single-site} entanglement where only one unit-cell $x_L=x_R$ exists in the untruncated region $[x_L,x_R]=[x_L,x_L]$. For a single unit-cell, the Fourier phase factor $\text{e}^{ik(x_1-x_2)}$ in $\langle {x,\alpha}| P|x,\beta\rangle$ (Eq.~\eqref{pp}) disappears, and the single unit-cell truncated projector (we call it $\tilde P$ here to emphasize the single unit cell special case) is
\begin{equation}\label{spp1}
	\begin{split}		
		\tilde P^{+-}&=-\displaystyle\frac{1}{{L}}\sum\limits_k P^{+-}_K \xrightarrow{L\gg1} -\displaystyle\frac{1}{L} P^{+-}_{\pi+i\frac{\alpha}{L+1}} +\frac{1}{2\pi}\int_{k\neq \pi} P^{+-}_{k+i\frac{\alpha}{L+1}} \text{d}k\\
		&=\frac{1}{2L}\frac{\sqrt{t_L -  \text{e}^{\frac{\alpha}{L+1}}}}{\sqrt{t_R -  \text{e}^{-\frac{\alpha}{L+1}}}} +\frac{1}{2\pi}\int_{k\neq \pi} \frac{\sqrt{t_L -  \text{e}^{-ik+\frac{\alpha}{L+1}}}}{\sqrt{t_R -  \text{e}^{ik-\frac{\alpha}{L+1}}}}\text{d}k\\
		&\xrightarrow[t_R -  \exp(\frac{\alpha}{L_c-1}) = 0]{}\frac{\sqrt{t_Lt_R -1}}{2t_R L}\left(\frac{\alpha}{L+1}-\frac{\alpha}{L_c+1}\right)^{-1/2}-\frac{1}{\pi}\sqrt{\frac{-t_L}{t_R}}\mathcal{F}\left(\text{asin}\sqrt{-t_R \text{e}^{-\frac{\alpha}{L+1}}},\frac{1}{t_Rt_L}\right)\ ,
	\end{split}
\end{equation}
where $\mathcal{F}(m,\phi)$ represents the Elliptic integral of the second kind, which varies very weakly with system size $L$ due to the slow $e^{-\frac{\alpha}{L+1}}$ functional dependence, and can be approximated as $\sqrt{(t_L+1)(t_R+1)}/2\pi t_R$. Hence the above is approximated by 
\begin{equation}\begin{split}\label{spp0}
		\tilde P^{+-}&\approx\frac{\sqrt{t_Lt_R -1}}{2\sqrt{\alpha}\cdot t_R} (L-L_c)^{-1/2}+\frac{\sqrt{(t_L+1)(t_R+1)}}{2\pi t_R}\ ,\\
\end{split}\end{equation}
with $\sqrt{(L+1)(L_c+1)}\approx L$ near the entanglement dip. By contrast, the other off-diagonal term  {$\tilde P^{-+}$} 
is not sensitive to $L$, as given by
\begin{equation}\begin{split}\label{spp}
		\tilde P^{-+}&= -\displaystyle\frac{1}{{L}}\sum\limits_k P^{-+}_K\approx\frac{ \sqrt{(t_L+1)(t_R+1)}}{2\pi t_L}\ .\\
\end{split}\end{equation}

\begin{figure}[H]
	\centering
	\includegraphics[width=0.7\linewidth]{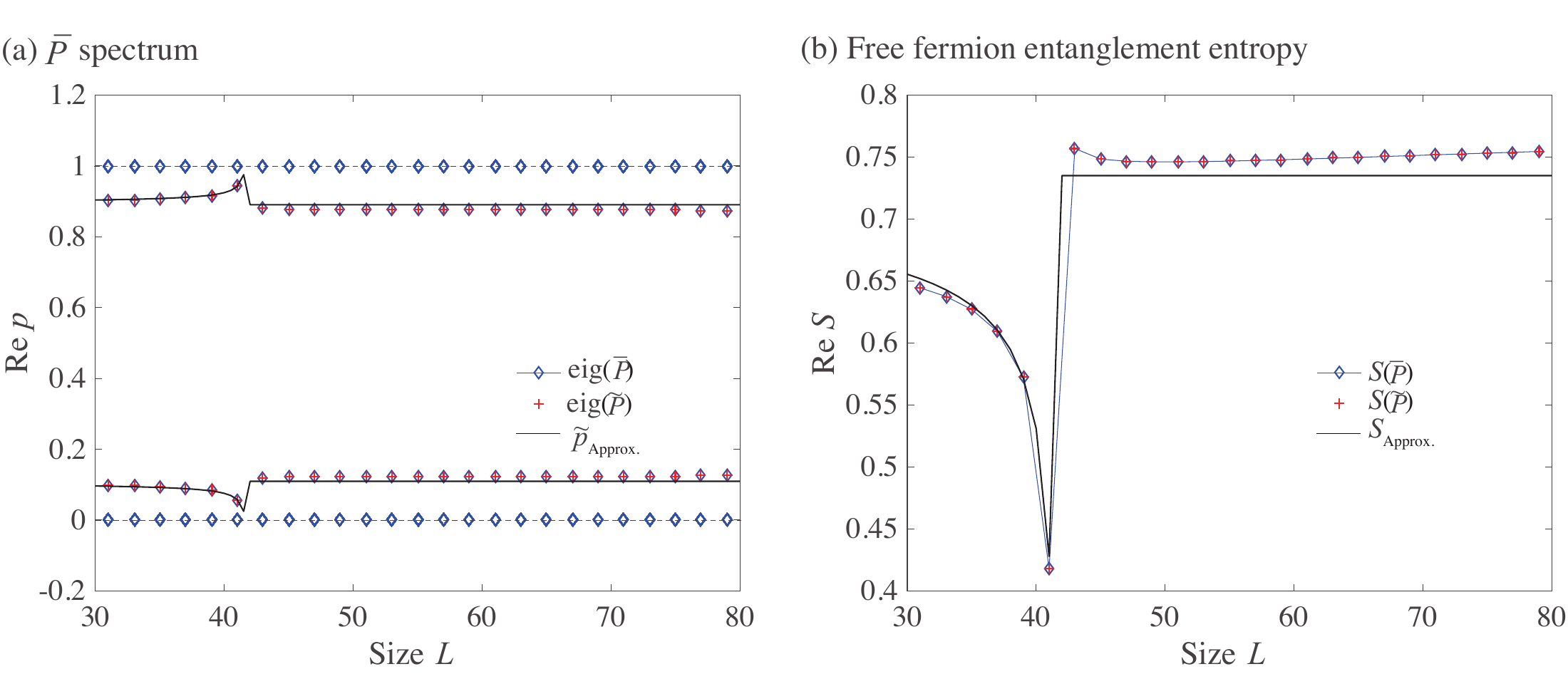}
	\caption{	
{\bf Scaling behavior of the $\bar{P}$ spectrum and the EE of the 2-component model (Eq.~\eqref{Eqeffectivemodel}) with single unit-cell truncation.} 
(a) The excellent agreement between the numerical spectra of $\bar{P}$ (Eq.~\eqref{2bandP3}) and its single unit-cell approximation $\tilde{P}$ (Eq.~\eqref{spp0}), as well as its analytically approximated spectrum $\tilde{p}_\text{Approx.}$ (Eq.~\eqref{spp1}). Note that $\tilde{P}$ only yields the nontrivial pair of eigenvalues away from 0 and 1, whose kink near $L\approx L_c$ is well-approximated by $\tilde{p}_\text{Approx.}$. 
	(b) The corresponding scaling of the entanglement entropy (EE) $S(\bar{P})$ and $S(\tilde{P})$ from the eigenvalues of
	$\bar{P}$ and $\tilde{P}$ shown in panel (a), along with the analytical result $S_\text{Approx.}$ given by Eq.~\eqref{sfit}. The results exhibit good consistency, and in particular all predict the same entanglement dip. 
	Parameters are $t_L = 1.2\text{e}^{0.3},t_R = 1.2\text{e}^{-0.3}, \delta=1.6\times 10^{-3}$. The coefficients $a$ and $b$ in Eq.~\eqref{sfit} are calculated as $a = 0.3905$ and $b=0.0456$ with small constants {$\mathcal{O}=0.14$, $\mathcal{O}'=-0.035$} (Eq.~(\ref{spp1},\ref{sfit})), corresponding to the critical scale being $L_c=42.79$.
	}
	\label{picS6}
\end{figure}

Hence the  {occupation} probability eigenvalues $\tilde p$ of $\tilde P$ are given by
\begin{small}
	\begin{equation}\label{spp1}
		\begin{split} 
			&\tilde{p}_{ {\text{Approx.}}} = \frac1{2}\pm \sqrt{\tilde{P}^{+-}\times\tilde{P}^{-+}}\\
			&=\frac1{2}\pm \left(\displaystyle\frac{1}{4}{\displaystyle\frac{t_R}{t_L}}\sqrt{\displaystyle\frac{t_L/t_R-1/t^2_R}{\alpha}} \frac{1}{\sqrt{L-L_c}}+  \frac{1}{2\pi\sqrt{t_Rt_L}}\sqrt{(t_L+1)(t_R+1)}\right)+\mathcal{O}\ ,
		\end{split} 
	\end{equation}
\end{small}
where the small constant $\mathcal{O}$ is of the order of $10^{-1}$. For entanglement dips that are not too deep (i.e $\tilde p\approx 0$ or equivalently $\tilde p\approx 1$), we can further approximate the EE by
\begin{equation}\label{sfit}
	\begin{split}
		S_{\text{Approx.}}
		&\xrightarrow{|\tilde p| \rightarrow 0 }-2\left(\tilde p\log\tilde p -(1-\tilde p)\tilde p\right)\\
		&=2(b\log(0.5-a)-2ab)(L_c-L)^{-0.5}+0.25+a^2+\mathcal{O}' )\ ,
	\end{split}
\end{equation}
where $a=\frac{1}{2\pi\sqrt{t_Rt_L}}\sqrt{(t_L+1)(t_R+1)}+ \mathcal{O}$, $b=\frac{1}{4}{\frac{t_R}{t_L}}\sqrt{\frac{t_L/t_R-1/t^2_R}{\alpha}}$,  
$\mathcal{O}$ and $\mathcal{O}'$ representing small offsets that can be adjusted to mitigate the imperfect approximation (See Fig.~\ref{picS6}b). 

\subsubsection{Effect of the length of the entanglement truncation interval}

\begin{figure}
	\centering
	\includegraphics[width=.8\linewidth]{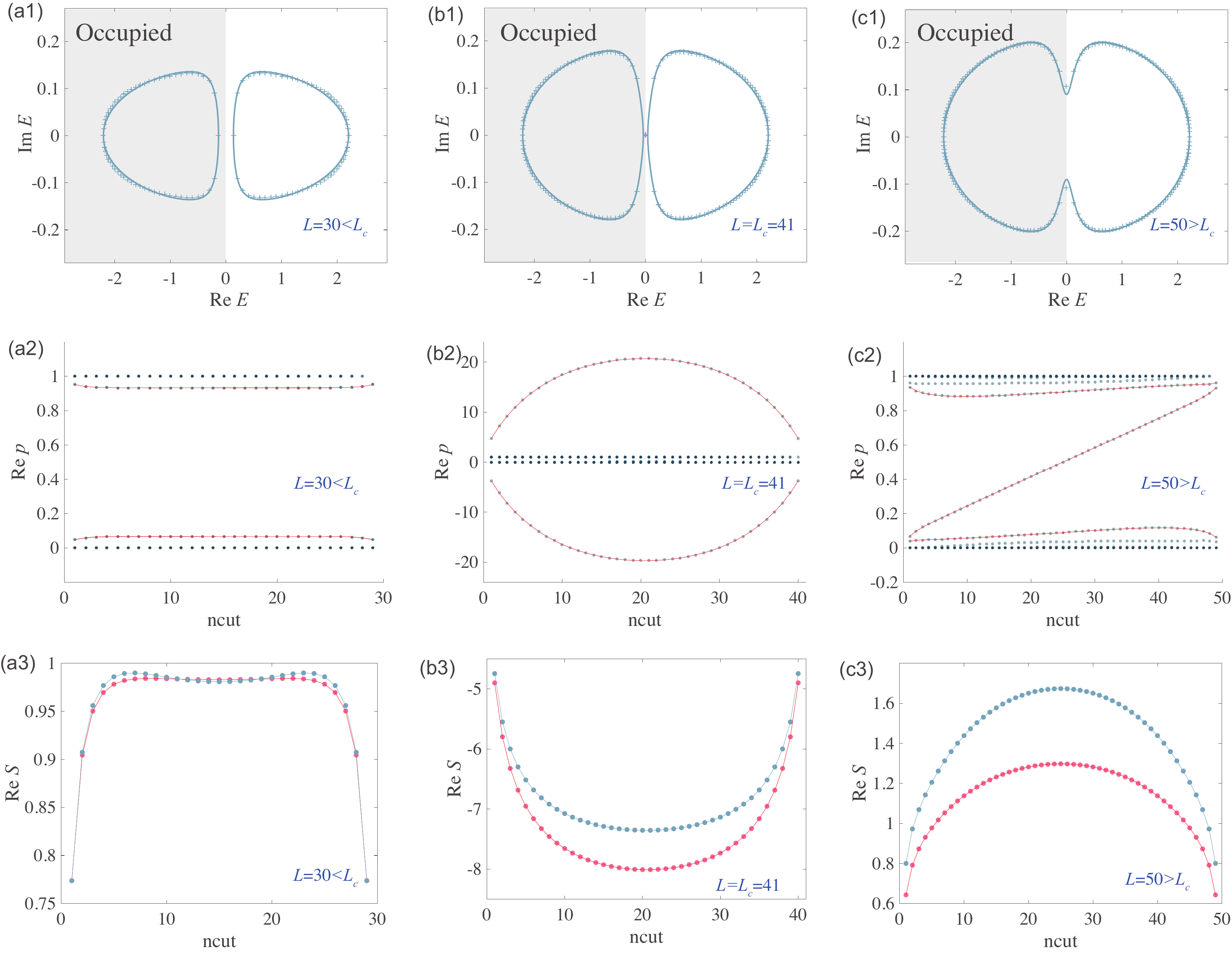}
	\caption{
	{\bf Scaling of the energy spectrum, $\bar{P}$ spectrum and entanglement entropy (EE) with different truncation lengths $n_{cut}$ in the 3 regimes: $L<L_c, L=L_c$ and $L>L_c$.}
	(a1-c1) The energy spectrum at different sizes $L$ where the Fermi energy is set at Re$E=0$, identifying the lower energy states in the gray area as the occupied states.
	(a2-c2) The $\bar{P}$ spectrum $p$ with the system divided into two partitions: the truncated region $x\in[1,n_{cut}]$ and non-truncated region $[n_{cut}+1, L]$ which is kept. Eigenvalues that are significantly far from 0 and 1 are joined by a red curve which exhibits qualitatively different behavior in different regimes.
	(a3-c3) Scaling of the EE with $n_{cut}$. For $L<L_c$, the total EE (blue curve) and the EE contributed only by the red states in (a2-c2) (red curve) are quite similar due to the suppression of the EE by the gap. But for the two other regimes, they differ considerably, albeit still with the same qualitative $n_{cut}$ dependence. Parameters used are $t_L = 1.2\text{e}^{0.3}, t_R = 1.2\text{e}^{-0.3}$, ${\delta=1.6844163\times10^{-3}}$.
	}	
	\label{picS10}
\end{figure}

In Hermitian systems, it is well-known that the EE of a cut region $[1,n_{cut}]$ varies strongly like $S\sim \log\left(L\sin \frac{\pi n_{cut}}{L}\right)$, as can be proven with boundary CFT. Hence it is prudent to check whether SIEC behavior is nontrivially influenced by $n_{cut}$. 

In Fig.~\ref{picS10}, it was found that the effect of $n_{cut}$ qualitatively depends on whether $L<L_c$, $L\approx L_c$ or $L> L_c$. In the first regime $L<L_c$ before the onset of the EE dip, the system is essentially gapped and $n_{cut}$ does not appreciably change either the $\bar P$ spectrum or $S$. However, for the second regime $L\approx L_c$ where the EE dip occurs, the $\bar P$ eigenvalues do become significantly enhanced at $n_{cut}\approx L/2$, lying very far out of $[0,1]$. For the third regime $L>L_c$, the system behaves essentially like an ordinary gapless system, with positive $S$ and $\bar P$ eigenvalues lying within $[0,1]$. One state traverses the $\bar P$ spectral gap between $0$ and $1$, reminiscent of the spectral flow of a topological edge mode.

Interestingly, as shown in the bottom row of Fig.~\ref{picS10}, despite the very different $\bar P$ behavior at or after the onset of the EE dip (second and third regimes), the EE continues to adhere approximately to the conventional $S\sim \log\left(L\sin \frac{\pi n_{cut}}{L}\right)$ behavior as $n_{cut}$ varies (for fixed $L$).

\subsection{Controlling the depth of entanglement dips}\label{suppmat2c}


Below, we explore the relationship between the depth of the entanglement dip (i.e. minumum EE $S_\text{min}$) varying the coupling parameter $\delta$ while keeping other parameters constant. This approach is based on our analytical understanding of how $L_c$ varies with $\delta$ (See Fig.~\ref{picS2}).  Here, we numerically identify parameters that lead to exceptionally negative $S_\text{min}$ values, where the entanglement dip represents drastic departures from usual $\log L$ entanglement scaling.

\begin{figure}[H]
	\centering
	\includegraphics[width=.7\linewidth]{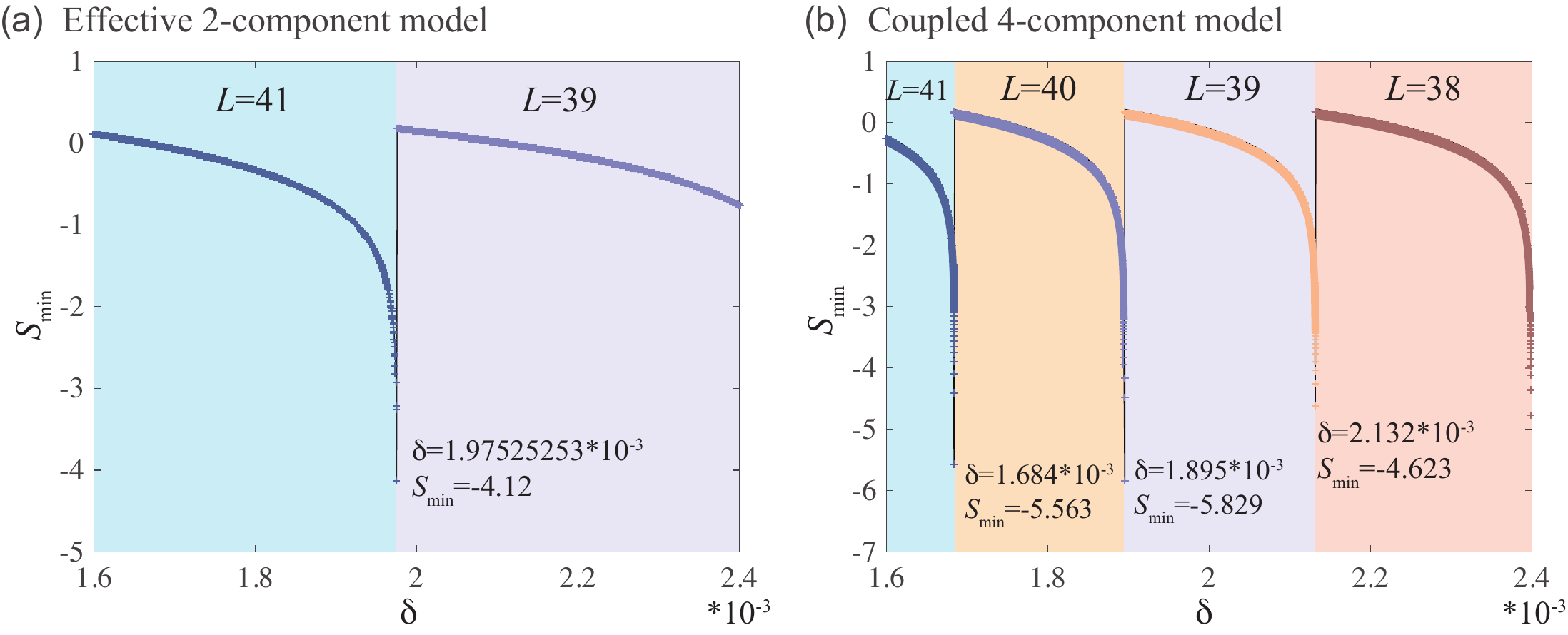}
	\caption{{\bf Sensitivity of the entanglement dip minimum $S_\text{min}$ with coupling $\delta$.}
The entanglement dip $S_\text{min}$ can yield negative EE for a wide range of parameters, and can even reach very negative values of $S_\text{min}<-4$ across small parameter windows. Shown are the $S_{\text{min}}$ and the corresponding system chain length $L$ where it is attained, both for our (a) 2-component effective model Eq.~\eqref{Eqeffectivemodel} with $\delta$ built into its scale-dependent GBZ and (b) its parent 4-component model Eq.~\eqref{EqS1} with $\delta$ being the physical inter-chain coupling. A more complete parameter space plot that shows the dependence of $S_\text{min}$ with $t_L/t_R$ is given in Fig.~3(c) of the main text. Parameters are $t_L = 1.2\text{e}^{0.3}, t_R = 1.2\text{e}^{-0.3}$. }	
	\label{picS7}
\end{figure}

Although the theoretically predicted $L_c$ varies continuously with $\delta$ (Eq.~\eqref{Eqcritical1}), an actual lattice contains only an integer number of unit cells $L$. This prevents us from getting infinitesimally close to $L_c$, where $S$ truly diverges. Away from that, how negative a dip $S$ can reach depends on commensurability considerations, as shown in Fig.~\ref{picS7}. For certain fine-tuned values of $\delta$, the EE through $S_\text{min}$ can dip below $-4$ (such as at $\delta$=1.68441635 $\times 10^{-3}$ or $2.131947\times 10^{-3}$), corresponding to extremely large $p$ eigenvalues. Such dips occur both for our effective two-component model (a) as well as its parent four-component model (b), even though the dip positions are different. But what remains consistent are the shapes and qualitative order of magnitude of the entanglement dips.




\subsection{Suppressed NHSE in the entanglement eigenstates }\label{suppmat2d}

Since $P$ is a projector onto the occupied bands, the entanglement eigenstates $\Psi_p$ with $\bar P\Psi_p =p\Psi_p$,  $\bar P=\bar{\mathcal{R}_{[x_1,x_2]}}P\bar{\mathcal{R}}_{[x_1,x_2]}$ are expected to be approximate linear combinations of the occupied states that vanish outside of the interval $[x_1,x_2]$. In particular, they are expected to assume an exponentially decaying spatial profile if the constituent occupied states are all NHSE skin-localized states. 

\begin{figure}[H]
	\centering
	\includegraphics[width=0.55\linewidth]{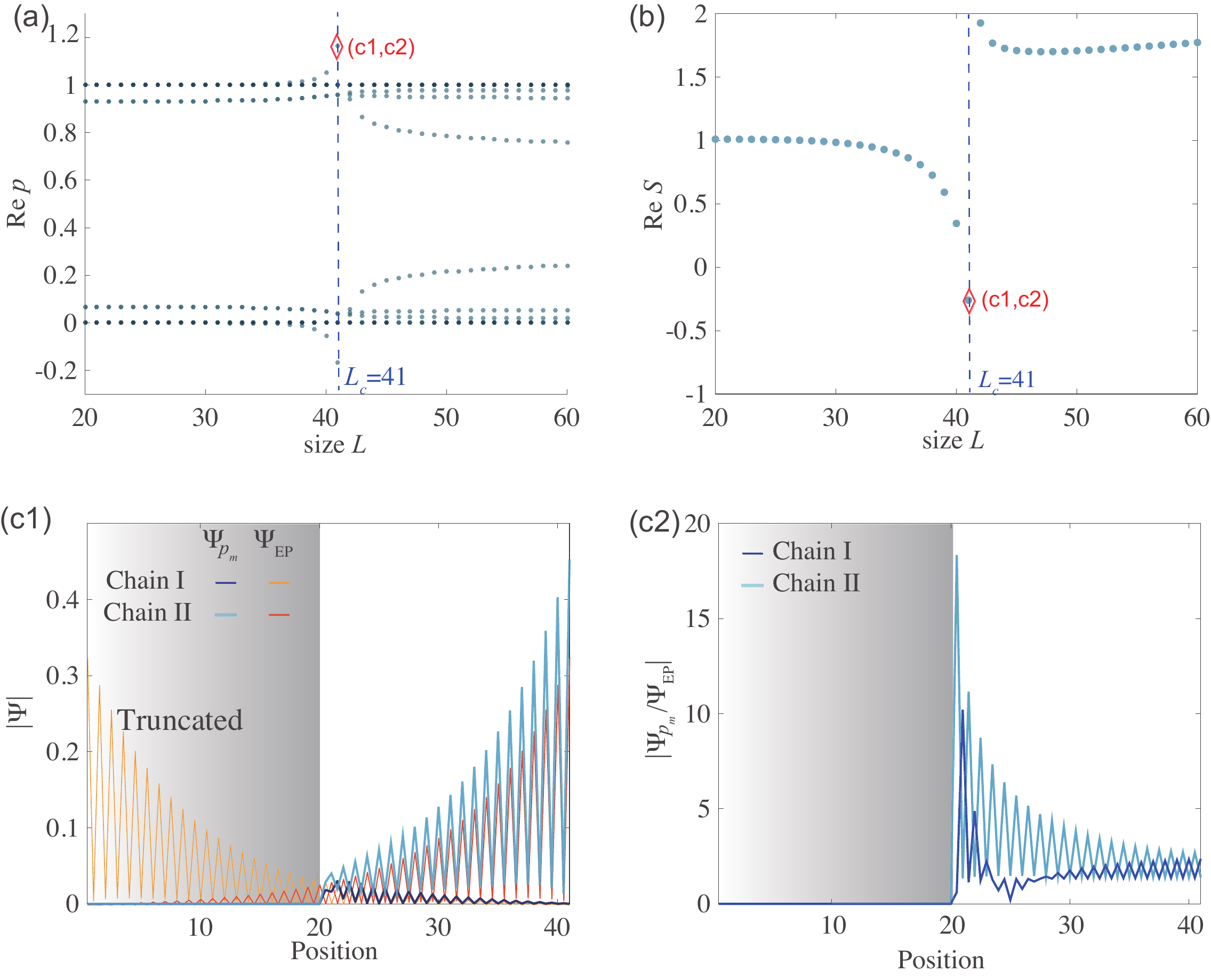}
	\caption{
	{\bf Entanglement scaling and $\bar P$ eigenstate profile for our physical 4-component model Eq.~\eqref{EqS1}.} 
	(a) $\bar P$ spectrum as a function of system size $L$ at half entanglement truncation. A conventional in-gap spectral branch for critical systems appears for $L>L_c$. 
	(b) The resultant entanglement entropy scaling, with a pronounced entanglement dip before $L\approx L_c$. 
	(c1) The spatial profiles of $\Psi_{p_m}$(the $S=S_\text{min}$ eigenstate of $\bar{P}$, colored red above) and $\Psi_\text{EP}$ (eigenstate of Hamiltonian $\mathcal{H}$) are shown. Since $\bar{P}$ is constructed using the eigenstates of the Hamiltonian, its eigenstate $\Psi_{p_m}$ inherits the skin effect from the Hamiltonian. (c2) To eliminate the skin effect in $\Psi_{p_m}$ and $\Psi_{\text{EP}}$, the profile of $|\Psi_{p_m}/\Psi_{\text{EP}}|$ represents the cumulative wavefunction distribution near the truncation point after excluding the EP eigenstate. Parameters used are $t_L = 1.2\text{e}^{0.3}, t_R = 1.2\text{e}^{-0.3}$, ${\delta=1.6\times10^{-3}}$.
	}
	\label{picS8}
\end{figure}

\begin{figure}[H]
	\centering
	\includegraphics[width=0.55\linewidth]{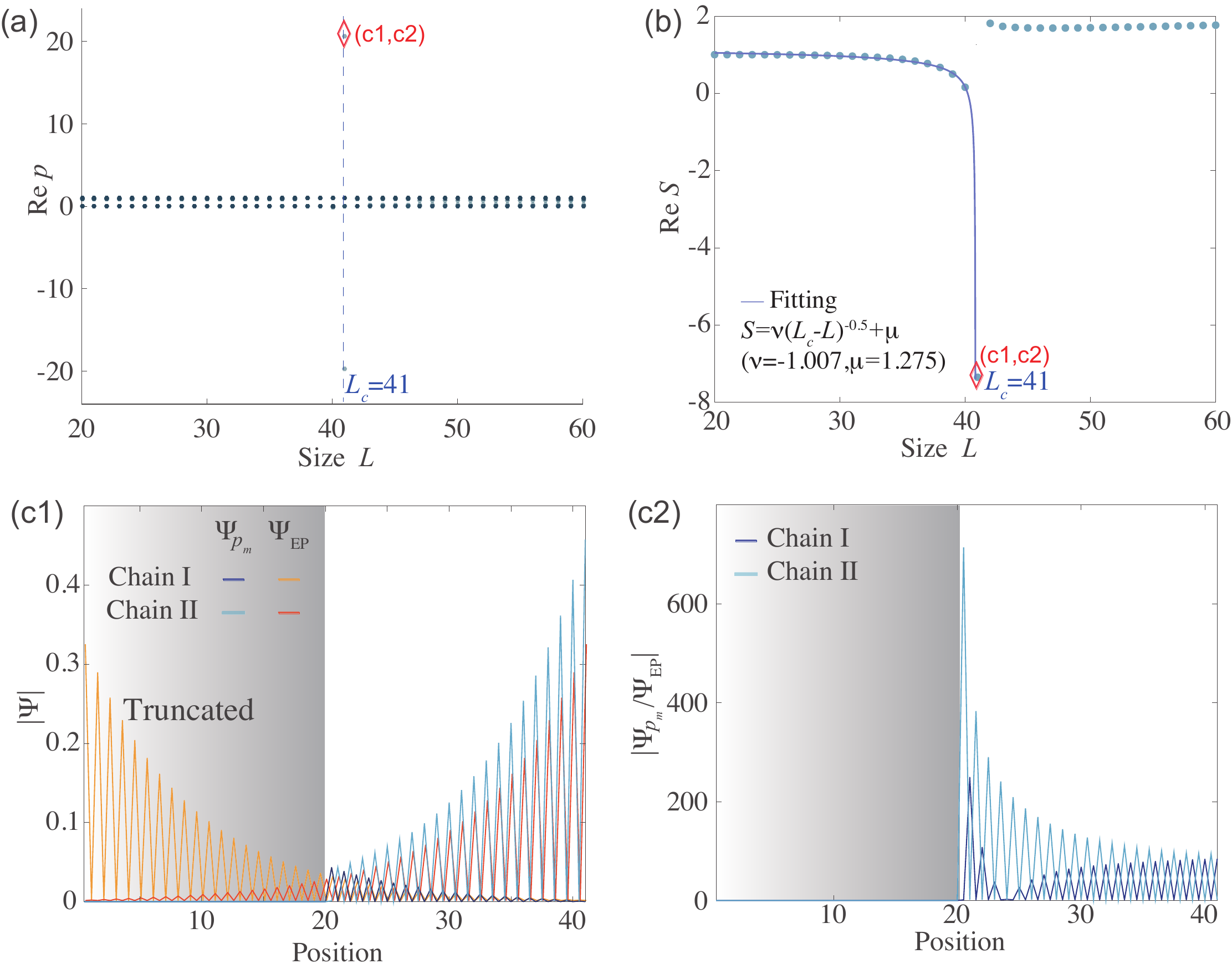}
	\caption{{\bf Entanglement scaling and $\bar P$ eigenstate profile for our physical 4-component model Eq.~\eqref{EqS1}, for an extremely pronounced entanglement dip.} This figure is similar to Fig.~\ref{picS8}, but with inter-chain coupling tuned to $\delta=1.6844163\times10^{-3}$ such that $L_c = 40.038$ becomes almost an integer, giving rise to very negative $S_\text{min}$. (a) and (b) shows how the $\bar{P}$ spectrum and corresponding entanglement entropy jump dramatically at $L\approx L_c$. 
However, as compared to Fig.~\ref{picS8}c1 and c2, the spatial distribution of $\Psi_{p_m}$ (depicted in (c1) and colored red in (a,b)) and $|\Psi_{p_m}/\Psi_{\text{EP}}|$ in (c2) do not exhibit obvious changes.  
	Parameters are $t_L = 1.2\text{e}^{0.3}, t_R = 1.2\text{e}^{-0.3}$.
}	
	\label{picS9}
\end{figure}

Yet, as we show below, the $\Psi_p$ entanglement eigenstates exhibit reduced NHSE localization compared to the reference skin localization at the scaling induced exceptional point (EP). This is demonstrated in both Figs.~\ref{picS8} and \ref{picS9}. In the physical Hamiltonian, chain I exhibits the skin effect towards the right, while for chain II towards the left. Indeed, this skin localization is observed in both the physical Hamiltonian eigenstate $\Psi_\text{EP}$, as well as the eigenstate $\Psi_{p_m}$ of $\bar P$ corresponding to the most negative eigenvalue $p_m$, which dominates the contribution to the entanglement dip.

But, as evident from Figs.~\ref{picS8}c and \ref{picS9}c, the entanglement eigenstate $\Psi_{p_m}$ shows a weaker skin localization than $\Psi_\text{EP}$, despite being approximately made up of skin modes. This can be interpreted as a signature of NHSE suppression in the entanglement eigenstate, and can be qualitatively understood as a consequence of the antagonistic competition between the oppositely directed NHSE chains.

\section{Comparison with other types of critical scenarios}\label{suppmat3}

\begin{figure}[H]
	\centering
	\includegraphics[width=.75\linewidth]{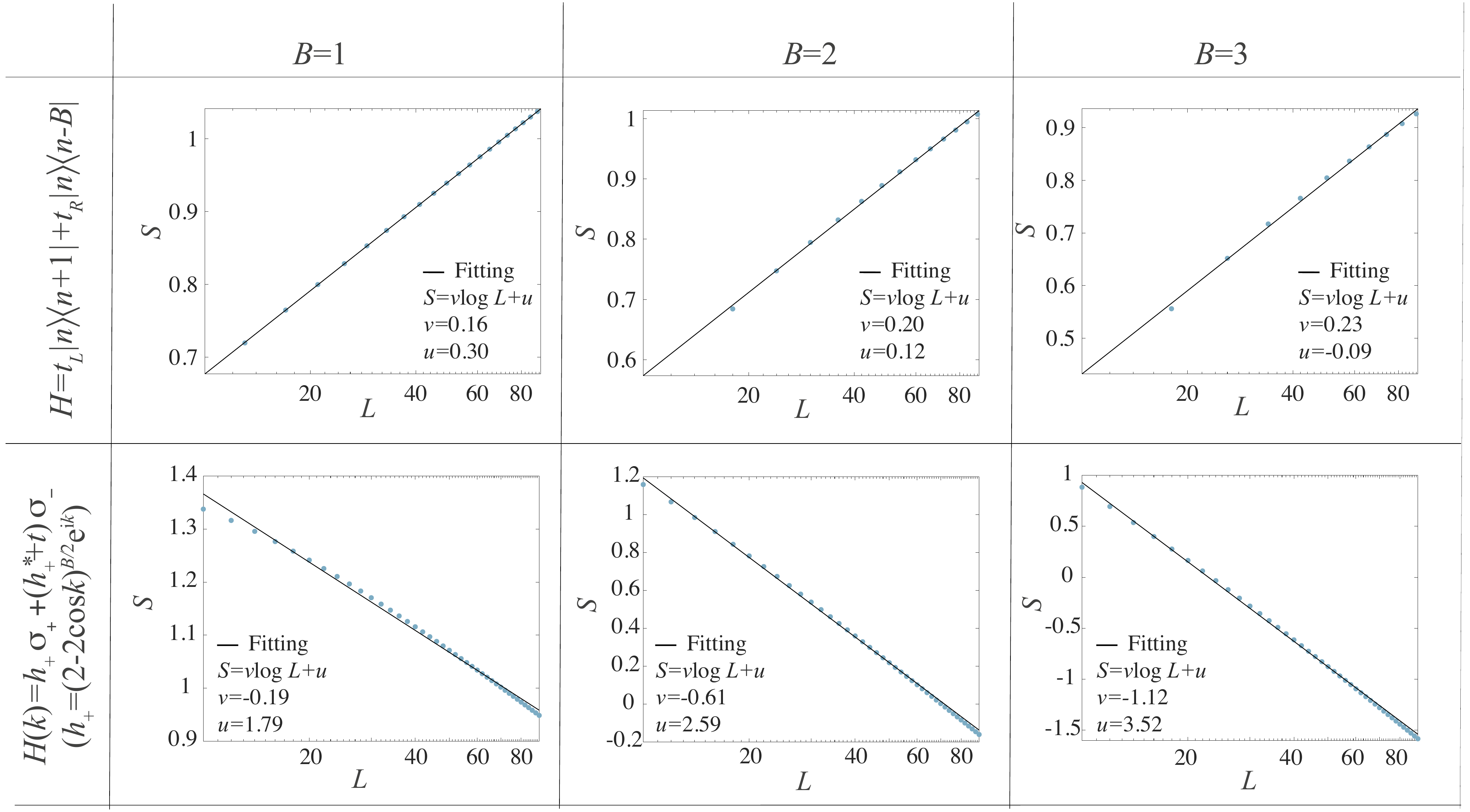}
	\caption{{\bf Logarithmic entanglement entropy scaling in generic non-SIEC systems that are critical.}
Shown are examples of the EE scaling of non-Hermitian systems exhibiting the NHSE (top) and exceptional points (bottom), labeled by maximal hopping distance $B$ as defined in the leftmost column. Regardless of form of the model, all cases exhibit $S \sim \log L$ scaling, even through $B>1$ gives rise to irregular generalized Brillouin zones (GBZs) or higher-order EP points. Only other systems exhibiting the SIEC can violate this logarithmic EE scaling. Parameters are $t_L=1.62,t_R=0.89, t=1$.}
	\label{picS12}
\end{figure}

\begin{figure}[H]
	\centering
	\includegraphics[width=1\linewidth]{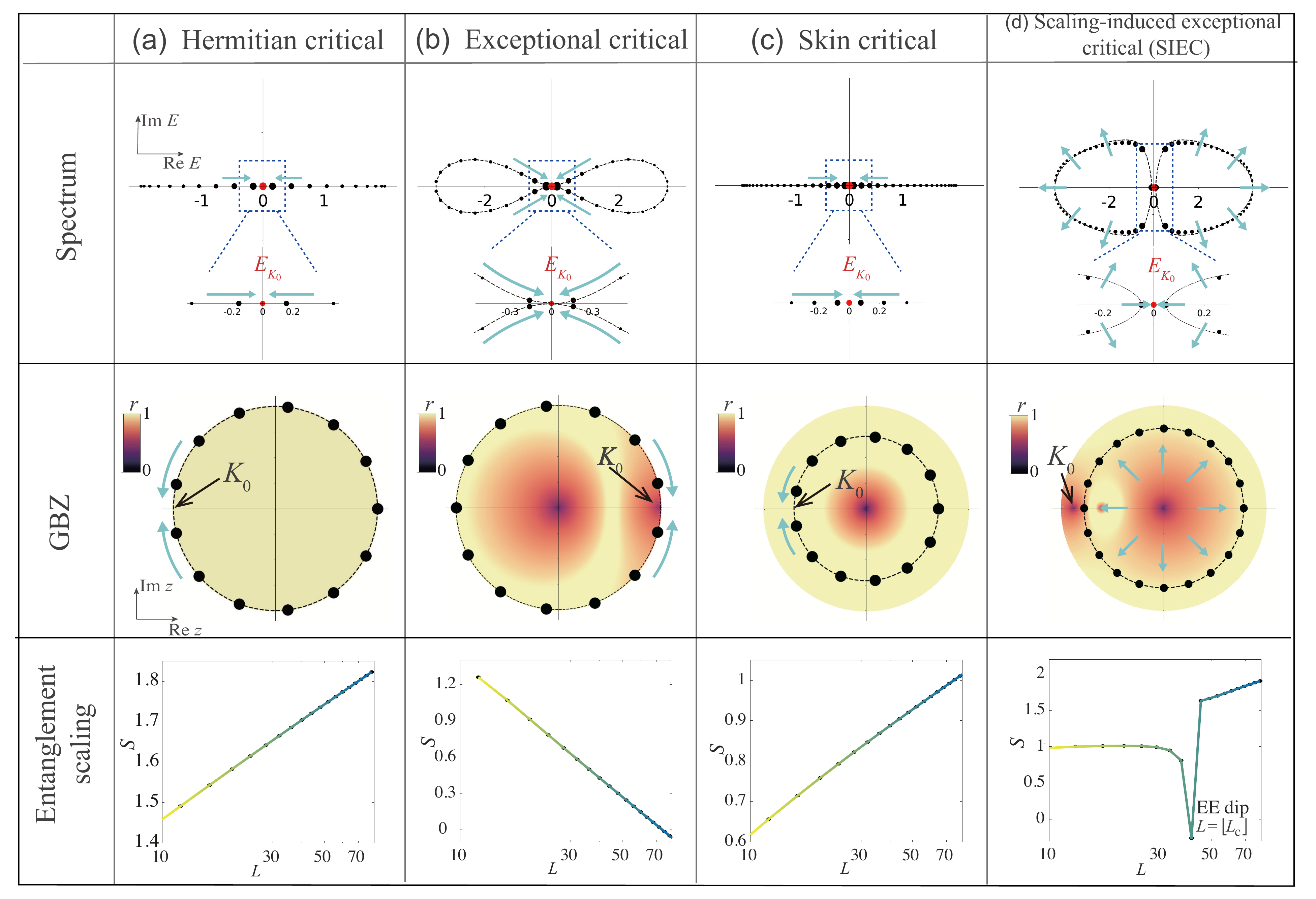}
	\caption{{\bf Entanglement scaling behavior of various types of critical systems, with $S\sim\log L$ scaling violated only in SIEC. } 
		(a) For a critical point $K_0$ that is Hermitian, the spectral accumulation of eigenmomentum states (black dots) intensifies around the corresponding energy $E_{K_0}$ (highlighted in red) with increasing system system size $L$ (cyan arrows). The nearest approach to $K_0$ hence scales like $\Delta K_\text{min}\sim \pi/L$, leading to logarithmic entanglement entropy scaling with $L$, i.e. $S \sim \log L$. 
		The Hermitian model used is $H(z) = \left( t_1 + t_2z\right)\sigma_- + \left( t_1 + t_2z^{-1}\right)\sigma_+, z = \text{e}^{ik},k\in \mathbb{R}, t_1=t_2=1.$ 
		(b) Even under the influence of the non-Hermitian skin effect (NHSE), the logarithmic scaling of entanglement entropy $S\sim \log L$ persists for a critical system because we still have $\Delta K_\text{min}\sim \pi/L$ in the generalized Brillouin zone (GBZ) $z=\text{e}^{iK (k)}$ (circular path within the unit circle).
		The non-Hermitian model used is $H(z) = \left( t_R + t_2z\right)\sigma_- + \left( t_L + t_2z^{-1}\right)\sigma_+,  t_L=t_2=1,t_R=0.5$. 
		(c) For an exceptional non-Hermitian critical point $K_0$, the eigenmomentum convergence $\Delta K_\text{min}$ remains proportional to $\pi/L$. Nevertheless, the logarithmic entanglement entropy $S \sim -\log L$ exhibits negative scaling, attributable to the divergence of 2-point functions due to geometric defectiveness (vanishing phase rigidity $r\rightarrow 0$ (dark purple)). The non-Hermitian exceptional point model used is $H(z) = \left( (2 - 2\cos k) z^{-1}\right)\sigma_- + \left( (2 - 2\cos k )z + 1\right)\sigma_+, z = \text{e}^{ik},k\in \mathbb{R}.$ 
		(d) In the case of scaling-induced exceptional criticality (SIEC), both the spectrum and GBZ changes dramatically with system size $L$ (visualized by widening cyan arrows), encountering the exceptional critical point $K_0$ only at a particular system size $L=L_c$. This condition induces a characteristic entanglement dip in entanglement entropy $S$, as explained in the previous sections. The SIEC model used is the same as that in Fig.~1 of the main text.
		\label{picS11}}
\end{figure}

To expand on the discussion surround Fig.~1 of the main text, we present in Fig.~\ref{picS12} other examples of NHSE and geometrically defective systems that all adhere to logarithmic entanglement scaling, and in Fig.~\ref{picS11} an elaboration of all the cases featured in Fig.~1 of the main text. 

\section{ Generality of entanglement dips} SIEC and entanglement dips are not limited to the coupled $H_\text{SSH}$ chains explicitly computed so far. They are expected to show up whenever the spectrum depends on $L$, and becomes gapless at an EP at a special $L\approx L_c$. This requires a scale-dependent GBZ that can be generically designed by coupling subsystems with competing NHSE pumping. 
Presented in Fig.~\ref{fig:fig4} are other models containing oppositely-directed NHSE channels, albeit with lower symmetry. Evidently, they all exhibit scale-dependent GBZs, and importantly exhibit entanglement dips at $L\approx L_c$ whenever the GBZ encounters a critical point $K_0$ (red). 

\begin{figure}
	\centering
	\includegraphics[width=0.6\linewidth]{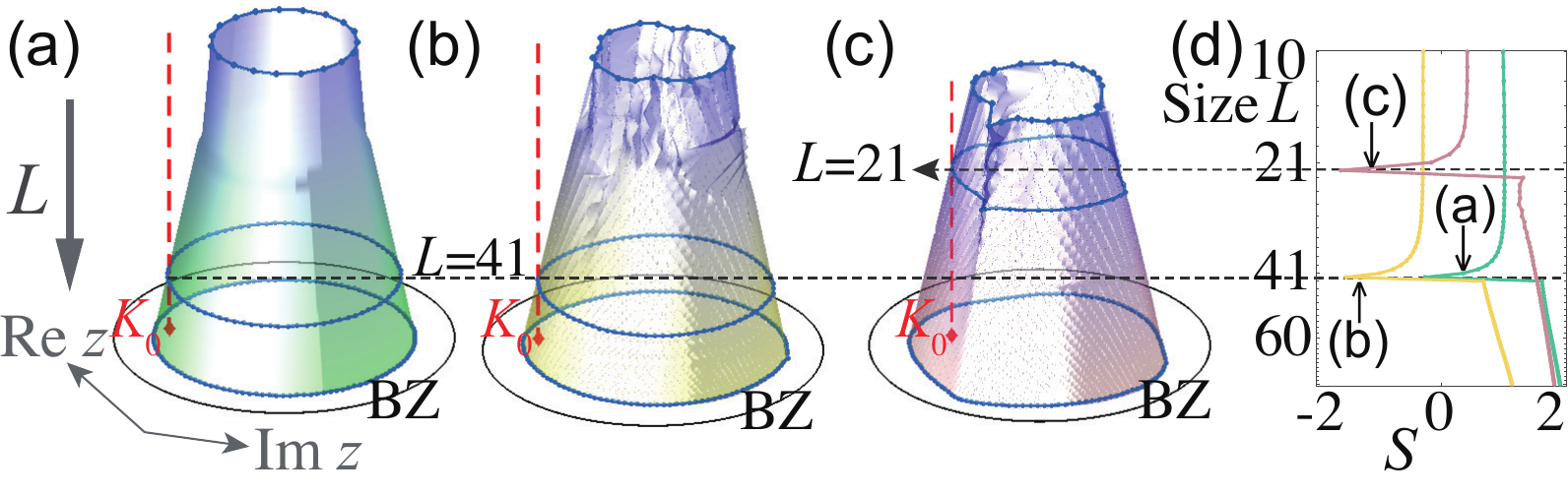}
	\caption{
Entanglement dips and scale-dependent GBZs in generalized models satisfying Eq.~\eqref{EqS1}.
(a–c) $L$-dependent generalized Brillouin zones (GBZs) for 3 illustrative models: 
(a) $H = (t_L + 1/z)\sigma_+ + (t_R + z)\sigma_-$, 
(b) $H = (t_L + \frac{1}{3}z^{-2})\sigma_+ + (t_R + z)\sigma_-$, and 
(c) $H = t_L\sigma_+ + (0.6 + 0.12i + z + 0.1z^{-2})\sigma_-$.
All models use weak inter-chain coupling $\delta = 1.6 \times 10^{-3}$ and asymmetric hopping $t_L, t_R = 1.2e^{\pm 0.3}$. 
For each case, GBZs are shown as a function of system size $L$, increasing from top to bottom. 
(d) Entanglement entropy $S$ exhibits clear dips at critical system sizes where the GBZ loops intersect the special momentum point $K_0$ (highlighted in red), despite the GBZs having irregular shapes.
}

\label{fig:fig4}
\end{figure}

\section{ Measuring Negative biorthogonal entanglement}\label{suppmat4}

A distinctive feature of non-Hermitian systems is that the left and right eigenstates are essentially eigenstates of \emph{different} Hamiltonians $\mathcal{H}$ and $\mathcal{H}^\dagger$. Together, these left and right eigenstates form a biorthogonal basis that preserves the probabilistic interpretation of quantum mechanics, and are crucial for defining various physical quantities such as our biorthogonal $\bar P$ and entanglement entropy $S$.

However, simultaneously obtaining information from both left and right eigenstates in experimental measurements is often challenging. Here, to address this difficulty, we suggest considering a larger system $\tilde{\mathcal{H}}$ comprising the original system $\mathcal{H}$ and its conjugate $\mathcal{H}^\dagger$. 
These two subsystems are very weakly coupled via a end-to-end coupling~\cite{torma2016physics,vale2021spectroscopic} $\bm \eta =\eta|1\rangle \langle L|+h.c.$ %
that connects the $|1\rangle$ and $|L\rangle$ end sites:
\begin{equation}\label{expri1}
	\begin{split}
		\tilde{\mathcal{H}}=  \begin{pmatrix}\mathcal{H}&\mathbf{\bm \eta} \\ \mathbf{\bm \eta^\dagger} &\mathcal{H}^{{\dagger}} \end{pmatrix}.
	\end{split}
\end{equation}
Importantly, this $\bm \eta$ coupling will not measurably affect the individual subsystems $\mathcal{H}$ and $\mathcal{H}^\dagger$'s eigenenergies and eigenstates, since they already contain equal and opposite NHSE themselves, and hence are no longer subject to net antagonistic NHSE. The purpose of these minuscule couplings $\eta$ is to enable cooperative response between the auxiliary system $\mathcal{H}^{\dagger}$ and the target system $\mathcal{H}$, and their impact on their fundamental characteristics, such as eigenergies and eigenstates, can be considered negligible.\\

The purpose of introducing the auxiliary Hamiltonian as $\mathcal{H}^{{\dagger}}$ is such that the eigenstates of the larger constructed Hamiltonian contains both the left and right eigenvectors of the target Hamiltonian. Specifically, for an energy $E$, there are 2 linearly independent degenerate eigenstates $|\tilde{\Psi}_{E,\pm}\rangle$ satisfying
\begin{equation}
	\begin{split}
		\tilde{\mathcal{H}}  |\tilde{\Psi}_{E,\pm}\rangle = E |\tilde{\Psi}_{E,\pm}\rangle,\qquad |\tilde{\Psi}_{E,\pm}\rangle=\begin{pmatrix}|\Psi^R_E\rangle \\ \pm K |\Psi^L_{E}\rangle  \end{pmatrix},
	\end{split}
\end{equation}
where $K$ is complex conjugation operation and $|\Psi^R_E\rangle$ and $|\Psi^L_E\rangle$ represent the eigenstates of the Hamiltonian $\mathcal{H}$ and its hermitian 
conjugate  $\mathcal{H}^{\dagger}$ ( $\mathcal{H}^{\dagger}= \mathcal{H}^{\text{T}}$) also with eigenvalue $E$.

\begin{figure}[H]
\centering
\includegraphics[width=0.4\linewidth]{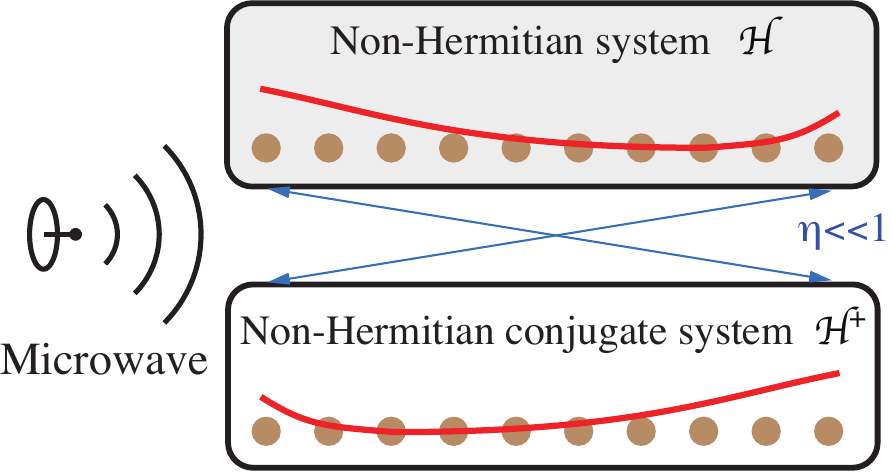}
\caption{ \label{picS13} (a)  A schematic setup in which the non-Hermitian system interacts with the auxiliary system through a weak microwave coupling. (b) A weak coupling $\eta$ are introduced between the endpoints of the $\mathcal{H}$ and $\mathcal{H}^{\dagger}$ systems. 
}
\end{figure}

Below, we show how to connect the biorthogonal normalization factor $\langle \Psi^L_{E} |\Psi^R_{E}\rangle$ of our target system $\mathcal{H}$ to the \emph{physically measurable} non-biorthogonal expectation in terms of the eigenstates of this enlarged Hamiltonian $\tilde{\mathcal{H}}$. We define an operator $\hat\Lambda=\sigma_-\otimes \mathbb{I}\,K$, which is the joint operation of $ \sigma_-\otimes \mathbb{I}$ and the complex conjugate operator $K$. The \emph{non-biorthogonal} i.e. conventionally measurable expectation of $\hat \Lambda$ in the energy $E$ eigenstate just gives the biorthogonal normalization factor:
\begin{equation}
\begin{split}
	\langle \hat \Lambda \rangle_E=\langle \tilde{\Psi}_{E,\pm} |\hat\Lambda|\tilde{\Psi}_{E,\pm}\rangle={\langle \Psi^L_{E} |\Psi^R_{E}\rangle}.
\end{split}
\end{equation}

Next, we show how the use of $\hat\Lambda$ can connect the non-biorthogonal expectation of an arbitrary operator $\hat A$ with its biorthogonal expectation:
\begin{equation}
\begin{split}\label{S52}
	\sum_{E,\pm}\langle \hat A\hat\Lambda \rangle_E/\langle \hat\Lambda \rangle_E &=\sum_{E,\pm}\frac{1}{\langle \tilde{\Psi}_{E,\pm} |\hat\Lambda|\tilde{\Psi}_{E,\pm}\rangle} \langle\tilde{\Psi}_{E,\pm}| \hat A\hat\Lambda |\tilde{\Psi}_{E,\pm}\rangle\\
	& = \sum_{E,\pm}  \frac{1}{\langle \tilde{\Psi}_{E,\pm} |\hat\Lambda|\tilde{\Psi}_{E,\pm}\rangle} \begin{pmatrix}\langle\Psi^R_E|  & \pm \langle\Psi^L_{E}| K\end{pmatrix} \begin{pmatrix} \bf 0& \bf 0\\ \hat A & \bf 0\end{pmatrix}  K\begin{pmatrix}|\Psi^R_E\rangle \\ \pm K |\Psi^L_{E}\rangle  \end{pmatrix}\\
	& \xrightarrow{K^2=\bf 1} \sum_{E,\pm}  \frac{1}{\langle \tilde{\Psi}_{E,\pm} |\hat\Lambda|\tilde{\Psi}_{E,\pm}\rangle} \begin{pmatrix}\langle\Psi^R_E|  & \pm \langle\Psi^L_{E}| K\end{pmatrix} \begin{pmatrix} \bf 0& \bf 0\\ \hat A & \bf 0\end{pmatrix}  \begin{pmatrix}K |\Psi^R_E\rangle \\ \pm |\Psi^L_{E}\rangle  \end{pmatrix}\\
	&= \sum_{E,\pm}  \frac{\pm 1 }{\langle \Psi^L_{E} |\Psi^R_{E}\rangle} \begin{pmatrix}\pm \langle\Psi^L_{E}| K\hat A & \bf 0\end{pmatrix}  \begin{pmatrix}K |\Psi^R_E\rangle \\ \pm |\Psi^L_{E}\rangle  \end{pmatrix}\\
	&=\sum_{E,\pm} \frac{ 1 }{\langle \Psi^L_{E} |\Psi^R_{E}\rangle}  \langle\Psi^L_{E}| K\hat A K\Psi^R_E\rangle\\
	&\xrightarrow{K\hat A=\hat A K}\sum_{E} \frac{ 2 }{\langle \Psi^L_{E} |\Psi^R_{E}\rangle}  \langle\Psi^L_{E}| \hat A|\Psi^R_E\rangle\\
	&=2\text{Tr} (\hat{A}P) = 2\langle A\rangle_{P}\ ,
\end{split}
\end{equation}
where $P=\sum\limits_E\displaystyle\frac{1}{\langle\Psi^L_E|\Psi^R_E\rangle} |\Psi^R_E\rangle \langle\Psi^L_E| $ is the biorthogonal projection operator, summed over the occupied energies $E$. Here, the factor of $2$ arises from the doubled  number of ($\pm$) states in $\tilde{\mathcal{H}}$. Although the biorthogonal expectation is arguably not directly measurable when we only have the target system $\mathcal{H}$, the above shows how we can express it in terms of the \emph{physical} expectations $\langle \hat A\hat\Lambda \rangle_E$ and $\langle\hat\Lambda \rangle_E$. In particular, by setting $\hat A=c_{i\alpha}^{\dagger} c_{j\beta} $ , we can obtain the two-point correlation functions $2\text{Tr} (P c_{i\alpha}^{\dagger} c_{j\beta})$, where $i$ and $j$ represent the unit cells and $\alpha$ and $\beta$ denote sub-lattices.\\
 
The 2nd Renyi entropy or purity can be measured based on the expectation values concerning the two-point correlation functions, or equivalently, the distribution of the density matrix. 
By evaluating the two-point correlation functions within an entanglement cut region, one can construct the truncated occupied state projector $\bar P$ of the subsystem. The 2nd Renyi entropy \( S_2 \) is then given by
$S_2 = -\log \text{Tr}(\bar P^2)$.
Therefore, by accurately measuring the expectation values corresponding to the correlation functions, we can derive the necessary information to compute the 2nd Renyi entropy.

Below, we provide a brief discussion of how some of the above steps may be implemented in quantum simulator systems. In principle, they are not restricted to any particular platform, but below we shall elaborate mostly in the context of ultracold atomic setups~\cite{li2019observation,ren2022chiral,lapp2019engineering,gou2020tunable,takasu2020pt,ferri2021emerging,rosa2022observing,qin2023non,shen2023proposal}, rather than trapped ions or solid-state spin systems.\\

In cold atom systems, employing either bosonic atoms, such as 87Rb, or fermionic atoms, such as 173Yb or 40K,  radio-frequency (RF) spectroscopy is a widely employed spectral measurement technique~\cite{torma2016physics,vale2021spectroscopic}.  It is based on weakly coupling the system to auxiliary energy levels using radio waves or microwaves, with weak coupling strength achievable by adjusting the laser power ~\cite{cohen1992laser,chu1991laser,muldoon2012control}. General RF spectroscopy as described by~\cite{li2022non,cao2023probing} can measure both the real and imaginary parts of eigenvalues of non-Hermitian systems.
And by inducing momentum transfer with coherent Raman laser beams and combining this with Time-of-Flight and absorption imaging techniques, the momentum distribution and spatial wave function of atoms can be accurately measured~\cite{metcalf1999laser}. For measuring expectation values, there are various techniques specifically geared towards cold atomic experiments~\cite{metcalf1999laser,ketterle1996evaporative,lett1988observation,denschlag2000generating}, such as time-of-flight measurements, absorption spectroscopy, Raman spectroscopy, and interference techniques.

Additionally, from an application perspective, active photonic systems have emerged as promising platforms for realizing non-Hermitian effects, including the skin effect studied here. In particular, hybrid photonic setups such as exciton-polariton condensates have demonstrated controllable gain and loss mechanisms that naturally implement non-Hermitian Hamiltonians~\cite{mandal2020nonreciprocal,mandal2022topological}. These systems allow for precise engineering of boundary conditions and coupling strengths, enabling the observation of non-Hermitian skin effects and related topological phenomena in a highly tunable environment. Furthermore, the inherent nonlinearity and driven-dissipative nature of active photonic platforms offer unique opportunities for dynamic control and switching applications in photonic devices. Therefore, the measurement schemes and theoretical frameworks discussed here could be adapted to explore and harness these effects in active photonic devices, enhancing their potential for technological applications.

Moreover, the simulation of non-Hermitian systems on quantum computing platforms has achieved remarkable progress~\cite{kamakari2022,koh2023measurement,peng2020simulating,chertkov2023characterizing,Shen2023,Shen2024,koh2024realization,koh2022prl,okuma2022nonnormal,Lan_2024,PhysRevA.111.032211,xie2024effectnoisequantumcircuit,Zhao2023,zhang2024observationnonhermitiansupersonicmode}. 
Specifically, the biorthogonal measurements required for a non-Hermitian system $\mathcal{H}$
 can be reformulated as joint measurements of an auxiliary operator $\hat\Lambda$ and a physical observable $A$ on the right eigenstate of a larger system $\tilde{\mathcal{H}}$(Eq.~\eqref{expri1}), as shown in Eq.~\eqref{S52}. 
 In quantum computing experiments, even the measurement of non-unitary operators can be efficiently implemented using advanced quantum algorithms and error mitigation techniques.
 This approach not only significantly broadens the experimental toolkit for validating non-Hermitian phenomena but also provides a solid technical foundation for exploring and harnessing unconventional quantum mechanical effects.

\end{document}